\documentclass{emulateapj}

\newcommand{\ovi}{O$\;${\small\rm VI}\relax}
\newcommand{\civ}{C$\;${\small\rm IV}\relax}
\newcommand{\siiv}{Si$\;${\small\rm IV}\relax}

\newcommand{\cii}{C$\;${\small\rm II}\relax}
\newcommand{\alii}{Al$\;${\small\rm II}\relax}

\newcommand{\feii}{Fe$\;${\small\rm II}\relax}
\newcommand{\siii}{Si$\;${\small\rm II}\relax}
\newcommand{\siiii}{Si$\;${\small\rm III}\relax}
\newcommand{\oi}{O$\;${\small\rm I}\relax}
\newcommand{\none}{N$\;${\small\rm I}\relax}
\newcommand{\nii}{N$\;${\small\rm II}\relax}
\newcommand{\nv}{N$\;${\small\rm V}\relax}
\newcommand{\pii}{P$\;${\small\rm II}\relax}
\newcommand{\sii}{S$\;${\small\rm II}\relax}

\newcommand{\cthree}{C$\;${\small\rm III}\relax}

\newcommand{\stwo}{S$\;${\small\rm II}\relax}
\newcommand{\sthree}{S$\;${\small\rm III}\relax}

\newcommand{\htwo}{H$_2$}

\newcommand{\halpha}{H$\alpha$}

\newcommand{\HI}{H$\;${\small\rm I}\relax}

\newcommand{\kms}{km~s$^{-1}$\relax}
\newcommand{\vlsr}{$v_{\rm LSR}$\relax}

\newcommand{\percc}{cm$^{-3}$}

\newcommand{\fuse}{{\em FUSE}}
\newcommand{\stis}{{STIS}}

\newcommand{\rosat}{{\em ROSAT}}

\newcommand{\hst}{{\em HST}}
\newcommand{\iue}{{\em IUE}}
\newcommand{\calfuse}{{\tt CalFUSE}}
\newcommand{\calstis}{{\tt CALSTIS}}

\newcommand{\zngi}{ZNG~1}

\begin{document}

\slugcomment{\bf Accepted by the {\em ApJ}}


\title{The High Velocity Gas toward Messier 5: Tracing Feedback Flows in the Inner Galaxy}

\author{William F. Zech\altaffilmark{1},
Nicolas Lehner\altaffilmark{1},
J. Christopher Howk\altaffilmark{1},
W. Van Dyke Dixon\altaffilmark{2},
\& Thomas M. Brown\altaffilmark{3}}

\altaffiltext{1}{Department of Physics and Astronomy,
  University of Notre Dame, Notre Dame, IN, 46556}

\altaffiltext{2}{Department of Physics and Astronomy,
  The Johns Hopkins University, Baltimore, MD, 21218}

\altaffiltext{3}{Space Telescope Science Institute,
  Baltimore, MD, 21218}


\begin{abstract}
  
  We present {\em Far Ultraviolet Spectroscopic Explorer} (\fuse) and Space 
Telescope Imaging Spectrograph (\stis\ E140M) observations of the post-asymptotic 
giant branch star \zngi\ in the globular cluster Messier 5 
($l=3\fdg9, \, b=+47\fdg7; \ d=7.5 \ {\rm kpc}, z=+5.3 \ {\rm kpc}$).  High 
velocity absorption is seen in \civ, \siiv, \ovi, and lower ionization species 
at LSR velocities of $\sim -140$ and $\sim -110$ \kms.  We conclude that this 
gas is not circumstellar on the basis of photoionization models and path length 
arguments.  Thus, the high velocity gas along the \zngi\ sight line is the 
first evidence that highly-ionized HVCs can be found near the Galactic disk.  
We measure the metallicity of these HVCs to be $[\rm{O/H}]=+0.22\pm0.10$, the 
highest of any known HVC.  Given the clouds' metallicity and distance constraints, 
we conclude that these HVCs have a Galactic origin.  This sight line probes gas 
toward the inner Galaxy, and we discuss the possibility that these HVCs may be 
related to a Galactic nuclear wind or Galactic fountain circulation in the inner 
regions of the Milky Way.

\end{abstract}



\section{Introduction}
\label{sec:intro}
High-velocity clouds (HVCs) are identified because they are clearly inconsistent 
with participating in Galactic rotation.  In practice, the cut-off for selecting 
HVCs is that they have LSR velocities $|\mbox{\vlsr}| \ge 90$ \kms.  While their 
origins are not fully understood, there is growing evidence that 
the HVCs are important components of the on-going exchange of matter between the 
Milky Way and the surrounding intergalactic medium (IGM) and, therefore, important 
to our understanding of the formation and evolution of galaxies (see, e.g., the 
recent reviews by Wakker et al. 1997, 1999; Wakker 2004; Benjamin 2004; 
Richter 2006).  In this context, many models have been constructed for the 
HVCs.  Oort (1970) first suggested the HVCs were gaseous relics left over from 
the formation of the Milky Way, and more recent models have elaborated on this 
model of HVCs as the building blocks of galaxies and the Local Group (Blitz et al. 1999).  The 
discovery of disrupted stellar satellites in the halo of the Milky Way has led 
to suggestions that HVCs may be the gaseous remnants of such satellites accreted 
by the Galaxy (Ibata et al. 1994; Putman et al. 2004), and the Magellanic Stream 
has long been known as gas removed from the Magellanic Clouds (Putman et al. 2003).  
Various models (e.g., Bregman 1978, Houck \& Bregman 1990) have also discussed the 
HVCs in the context of the galactic fountain, in which gas is ejected from the 
Galactic disk by the combined effects of multiple, correlated supernova explosions 
(Shapiro \& Field 1976; Norman \& Ikeuchi 1989).  These models have distinct 
predictions for the distances and metallicities of the HVCs.  These are the two 
principal diagnostics for the origins of the HVCs.

Over the last decade, observations with the {\em Hubble Space Telescope} (\hst), 
the {\em Far Ultraviolet Spectroscopic Explorer} (\fuse), and some ground-based 
instruments have provided measurements of the metallicities and distance brackets 
for some of the major neutral HVC complexes.  The general picture that has emerged 
for these prominent neutral complexes (e.g., complexes A and C) is one in which 
they have an extragalactic origin given their low metallicities (e.g., complex C 
has a metallicity $\sim$15\% solar; Collins et al. 2007, see van Woerden \& Wakker 
2004 for a summary) and distances from the sun bracketed to be within 
$5 \la d \la 12$ kpc (e.g., Thom et al. 2006, 2007; Wakker et al. 2007a,b; van 
Woerden et al. 1999).  The Magellanic Stream is another example, thought to 
be at $d\sim50-75$ kpc with a metallicity $\sim25\%$ that of the sun (Gibson 
et al. 2000, Sembach et al. 2001).

Traditionally, HVCs were observed via optical or radio observations 
(e.g., M\"unch 1952; M\"unch \& Zirin 1961, Wakker 1991).  These clouds were 
therefore known as neutral entities.  However, UV observations from {\hst} and 
\fuse\, as well as ground-based H$\alpha$ observations (Tufte et al. 2002, 
Putman et al. 2003), have revealed a significant ionized component of the 
neutral complexes. Furthermore, such data have also allowed the discovery of 
a new class of HVCs without \HI\ emission, the highly-ionized HVCs (Sembach 
et al. 1999, 2003; Lehner et al. 2001, Lehner 2002; Collins et al. 2004, 2005; 
Ganguly et al. 2005; Fox et al. 2006).  These HVCs show absorption from the 
``high ions'' \ovi, \nv, \civ, and \siiv; these ions require large energies 
for their production, and \ovi, in particular, cannot be produced via 
photoionization within the Galaxy (e.g., Sembach et al. 2003), suggesting 
the presence of hot gas to provide for the ionization. Some of the highly-ionized 
high velocity absorption seen towards AGNs is associated with the outer 
layers of known \HI\ HVC complexes based on their 
coincidental velocities and proximity on the sky (e.g., gas associated with 
the outer edges of complexes A, C, and the Magellanic Stream; Sembach et al. 
2003, Fox et al. 2004, 2005).  However, others have no 
\HI\ 21-cm emission, and therefore are not part of 
large neutral complexes.  They do, however, show \HI\ and \cthree\ {\it absorption} 
counterparts, implying they have a multiphase structure (Fox et al. 2006 and 
references therein).

The origin of these highly-ionized HVCs are mostly unknown because neither their 
distances nor their metallicities are well known. Up to this work, there has 
been no report of highly-ionized HVCs observed in absorption against distant 
Milky Way stars (Zsarg\'o et al. 2003), the only way to directly constrain 
the distances to HVCs.  This, together with kinematic arguments (Nicastro et al. 2003), 
led Nicastro (2005) to conclude that these HVCs 
must be extragalactic since the highly-ionized HVCs are detected only toward 
AGNs and QSOs. However, early-type stars often have complicated spectra near 
the high ions, and absorption from a highly-ionized HVC could easily be lost 
in the complicated structure of the continuum, especially if the absorption 
is weak.  The metallicities of highly-ionized HVCs are also difficult to determine.  
Since they are defined by their ionization state, the ionization corrections 
required to determine the metallicities are often large and extremely model 
dependent.  For the majority of these clouds, the distance and metallicity 
estimates rely on photoionization models based on an assumed ionizing spectrum 
(e.g., Sembach et al. 1999, Collins et al. 2004, 2005, Ganguly et al. 2005, 
and others).  These studies have suggested the clouds are located in the 
distant reaches of the Milky Way (e.g., Collins et al. 2005, Fox et al. 2006) 
and have significantly sub-solar metallicites, typically $\sim20\%$ solar with 
a range of 4\% to 40\% solar.  Prior to the present work, there have only been 
three metallicity limits for highly-ionized HVCs using the columns of \oi\ and 
\HI\ (Fox et al. 2005, Ganguly et al. 2005), a comparison that does not require 
ionization corrections to derive the metallicity.  These upper limits are mostly 
crude, consistent with both solar and sub-solar abundances.

In view of the uncertainties in their properties, the highly-ionized HVCs are 
consistent with both an extragalactic or Galactic origin. In particular, they 
may trace a hot Galactic fountain (Fox et al. 2006) or, in cases where the sight 
line passes near the Galactic center, a nuclear wind (Keeney et al. 2006).  
Galactic nuclear winds are observed in external galaxies across the electromagnetic 
spectrum (Martin 1999; Strickland 2002; Heckman 2002; Veilleux 2002), and there 
is evidence that the Milky Way also has a Galactic nuclear wind (Bland-Hawthorn 
\& Cohen 2003). Analyses of X-ray observations toward the Galactic center have 
provided further evidence for an outflow from the central regions of the Galaxy 
(Almy et al. 2000; Sofue 2000; Yao \& Wang 2007). Such a Galactic outflow/wind 
should be detectable via UV high-ion absorption not only in the spectra of QSOs 
but also in distant stars at high galactic latitudes.  Keeney et al. (2006) 
discussed the possibility that a wind from the Galactic center may give rise 
to highly-ionized HVC absorption seen in the spectra of two AGNs.  The subsolar 
 metallicities they attribute to these absorbers are at odds with a Galactic 
wind origin.  However, if, as they suggest, their photoionization models provide 
inappropriate metallicity estimates, there may yet be reason to believe these HVCs 
probe the expulsion (and perhaps subsequent return) of matter from the inner 
regions of the Milky Way.

In this work, we present new observations of a highly-ionized HVC that is located 
toward the inner Galaxy at a distance from the sun $d < 7.5$ kpc and height 
above the Galactic 
plane $z < 5.3$ kpc.  This HVC is detected in the \fuse\ and Space Telescope 
Imaging Spectrograph (\stis) spectra of the post-asymptotic giant branch (PAGB) 
star \zngi\ located in the globular cluster Messier~5 (NGC\,5904, $d = 7.5$ kpc).  
This is the first highly-ionized HVC for which the upper limit on the distance is 
known. This HVC was first reported by Dixon et al. (2004), but its origin 
(circumstellar or truly interstellar gas) was not fully explored in their work.  
Here we rule out the circumstellar origin and show the gas has a supersolar 
metallicity, which is consistent with an origin in the inner Galaxy.  This is 
the first highly-ionized HVC unambiguously detected in the spectrum of a Galactic 
star, and this HVC has the highest measured metallicity of any yet studied.

This paper is structured as follows.  We summarize the  \fuse\ and \stis\ 
observations and our reduction of the data in \S~\ref{sec:observations}, and 
in \S~\ref{sec:sightline} we discuss the properties of the star and sight line.  
We present our measurements of the observed high velocity absorption in 
\S~\ref{sec:analysis}. In \S~\ref{sec:physicalconditionsabundances} we discuss 
the physical properties of the high velocity gas including the electron density, 
ionization fraction, and metallicity. We analyze the kinematics and ionization 
mechanisms involved with the gas along the \zngi\ sight line in 
\S~\ref{sec:kinematics}.  We use photoionization modeling and path length 
arguments to rule out the circumstellar hypothesis for the origin of the 
high velocity gas toward \zngi\ in \S~\ref{sec:photoionizationmodel}. In 
\S~\ref{sec:discussion} we discuss our results in the context of a Galactic 
circulation/feedback and suggest further research to test this hypothesis, and 
we discuss the implications for other highly-ionized HVCs. We summarize our 
principal conclusions in \S~\ref{sec:summary}.


\section{Observations and Reductions}
\label{sec:observations}
 
The observations for this work were taken from \stis\ on board \hst\ and \fuse.  In 
the following subsections, we discuss the data reduction and handling procedures 
for the \stis\ and \fuse\ spectral data sets.


\subsection{\stis}
\label{sec:stis_observations}

The \stis\ observations were made between July 8 and July 19, 2003 under the Guest 
Observer program 9410.  Five visits were made with identifications O6N40401--402, 
and O6N40301--303 for a total exposure time of 12.8 ks.  The \stis\ observations 
of \zngi\ were taken in the ACCUM mode with the $0\farcs2 \times 0\farcs2$ aperture 
using the E140M echelle grating to disperse the light onto the far-ultraviolet 
Multi-Anode Microchannel-Array (MAMA) detector.  The usable wavelength coverage 
is from $\sim1150$ \AA\ to 1710 \AA.  The resolution of this mode is 
$R\equiv\lambda/\Delta\lambda\sim45,800$ corresponding to a velocity FWHM of 
$\sim$ 6.5 \kms\ with a detector pixel size of 3.22 \kms.  The \stis\ data were 
retrieved from the Multimission Archive at Space Telescope (MAST), and reduced 
with the \calstis\ (Version 2.14c; Brown et al. 2002) pipeline in order to 
provide orbital Doppler shift adjustments, detector nonlinearity corrections, 
dark image subtraction, flat field division, background subtraction, wavelength 
zero-point calibration, and to convert the wavelengths into the heliocentric 
reference frame.  The individual exposures were weighted by their inverse variance 
and combined into a single spectrum.  For a description of the design and 
construction of \stis\, see Woodgate et al. (1998), and a summary of the \stis\ 
on-orbit performance is given by Kimble et al. (1998).

We applied a shift of 
$\Delta\upsilon_{\rm{LSR}} = \upsilon_{\rm{LSR}} - \upsilon_{\rm{helio}} = +13.25$ 
\kms\ to the data to transform the heliocentric velocities provided by \stis\ 
to the local standard of rest (LSR) frame.  This assumes a solar motion of +20 
\kms\ in the direction 
$(\alpha,\delta)_{1900} = (18^h,+30^\circ)$ $[(l,b)\approx(56^\circ,+23^\circ)]$ 
 (Kerr \& Lynden-Bell 1986).  
For comparison, the Mihalas \& Binney (1981) definition of the LSR gives a 
velocity shift of $\Delta\upsilon_{\rm{LSR}} = +11.76$ \kms.   The velocity 
uncertainty of the \stis\ observations is $\sim$ 1 \kms\ with occasional 
errors as large as $\sim$ 3 \kms\ (see the Appendix of Tripp et al. 2005).


\subsection{\fuse}
\label{sec:fuse_observations}
The \fuse\ observations were made under programs A108 and D157 on July 15, 
2000 and between April 11 -- 13, 2003 for a total of four visits.  The \fuse\ 
data sets are A1080303 and D1570301--303, and the total exposure time is 30.4 ks.  
The four \fuse\ observations of \zngi\ were taken using the LWRS 
$30\arcsec \times 30\arcsec$ apertures in the photon event (TTAG) mode.  The 
wavelength range of the data is 905 \AA\ to 1187 \AA\ with a resolution of 
$\sim$ 20--25 \kms\ and a binned output pixel size of 3.74 \kms.  The data 
were reduced using the \calfuse\ (Version 3.1.3) pipeline.  The \calfuse\ 
processing is described in Dixon et al. (2007), and the spectroscopic 
capabilities and early on-orbit performance of \fuse\ are described in Moos 
et al. (2000) and Sahnow et al.(2000), respectively.  

Data were obtained from the SiC1, SiC2, LiF1, and LiF2 channels.  For each 
\fuse\ segment, the intermediate data files produced by \calfuse\ were shifted 
to a common wavelength scale and combined into a single file.  Time segments 
exhibiting a low count rate, e.g., when the target fell near the edge of the 
spectrograph apertures, were excluded from further consideration.  The detector 
and scattered-light background were scaled and subtracted by \calfuse.  The 
\fuse\ relative wavelength is accurate to roughly $\pm5$ \kms\ but can vary 
by $10-15$ \kms\ over small wavelength intervals.  The absolute zero point of 
the wavelength scale for the individual absorption lines is uncertain and was 
determined using the well-calibrated \stis\ data.  Where possible, transitions 
from the same species were compared (e.g., \ion{Fe}{2} $\lambda\lambda 1055, 1063$ 
were aligned with \ion{Fe}{2} $\lambda\lambda  1608, 1611$).  Where this was 
not possible, we compared different ions with similar ionization potentials 
and similar absorption depths (e.g., \ion{Ar}{1} $\lambda\lambda 1055, 1066$, 
with \ion{N}{1} $\lambda\lambda 1199, 1200$).  For the ion \ion{O}{6} we aligned 
and coadded the data from the LiF1A, LiF2B, SiC1A, and SiC2B channels.  No ion 
was available to directly fix the velocity scale of \ion{O}{6}.  However, the 
shifts for transitions throughout the LiF1A segment were all consistent, and 
we adopted their average to align \ion{O}{6}.  Based on a comparison of the 
velocity centroids of the \htwo\ lines $\lambda\lambda 1009, 1013, 1026, 1031$ 
near \ovi\ in \fuse, we conclude that no additional velocity shift for \ovi\ 
was needed.


\section{The M5 \zngi\ Sight Line}
\label{sec:sightline}

\begin{figure*}
\epsscale{1.0} 
\plotone{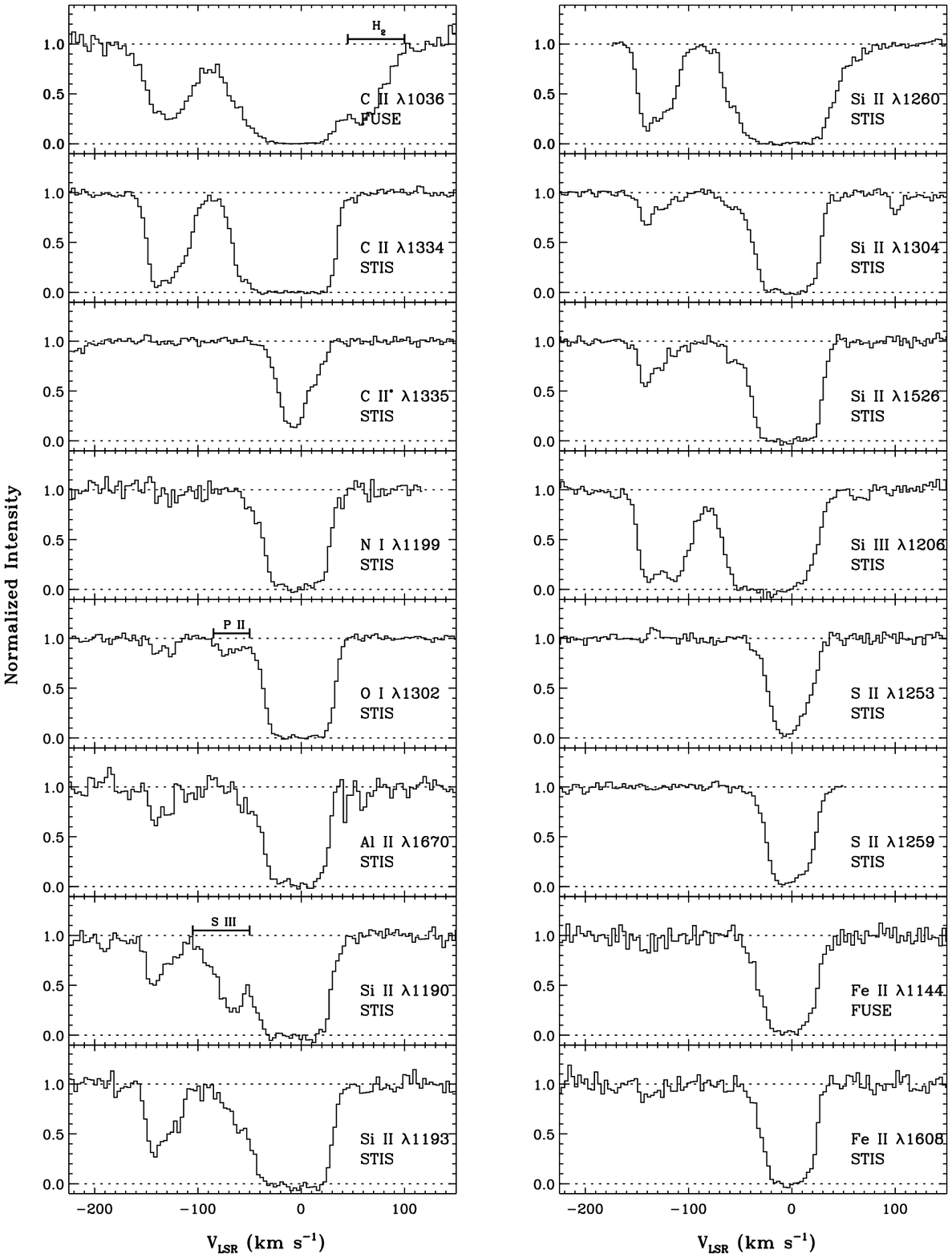}
\caption{Normalized intensity profiles of the tracers of the low ions.  
The HVC region lies in the velocity range $-160$ \kms\ to $-90$ \kms\ with 
centroid velocities for \civ\ and \siiv\ at $-142$ \kms\ and $-111$ \kms.  
Gas participating in 
the global rotation of the Galaxy is observed at $|\mbox{\vlsr}| \lesssim 30$ 
\kms.  The instrument from which the data were taken is identified in the lower 
right of each profile.  \fuse\ has a resolution of $\sim20$ \kms\ and \stis\ 
E140M has a resolution of $6.5$ \kms.
\label{f1}}
\end{figure*}

\begin{figure*}
\epsscale{1.0} 
\plotone{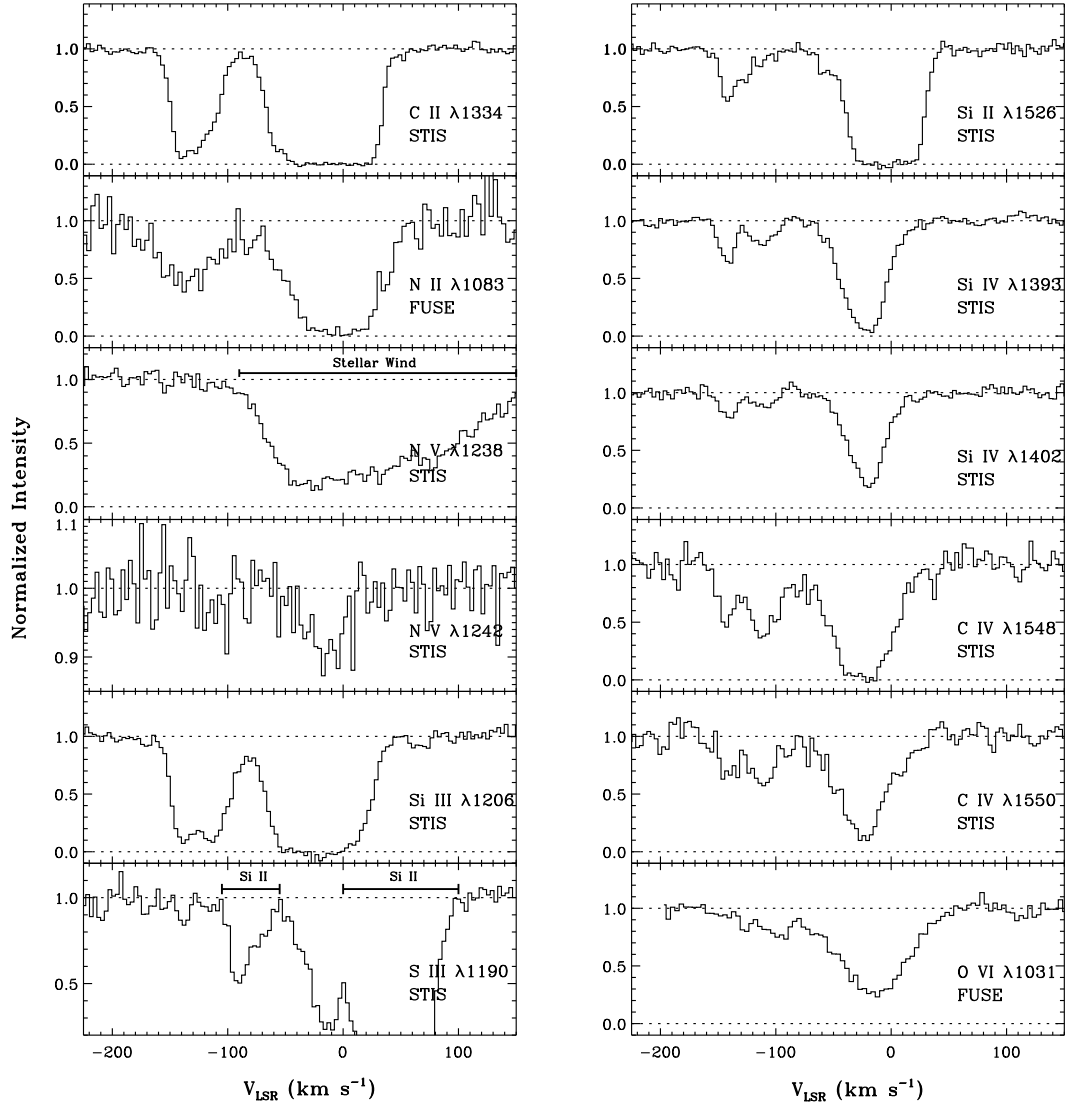}
\caption{Same as Figure~\ref{f1} but for the ions that are doubly or more 
ionized.  The \cii, \nii, and \siii\ profiles are for comparison (although 
note that \nii\ solely probes ionized gas).
\label{f2}}
\end{figure*}

\begin{figure}  
\epsscale{1.2}
\plotone{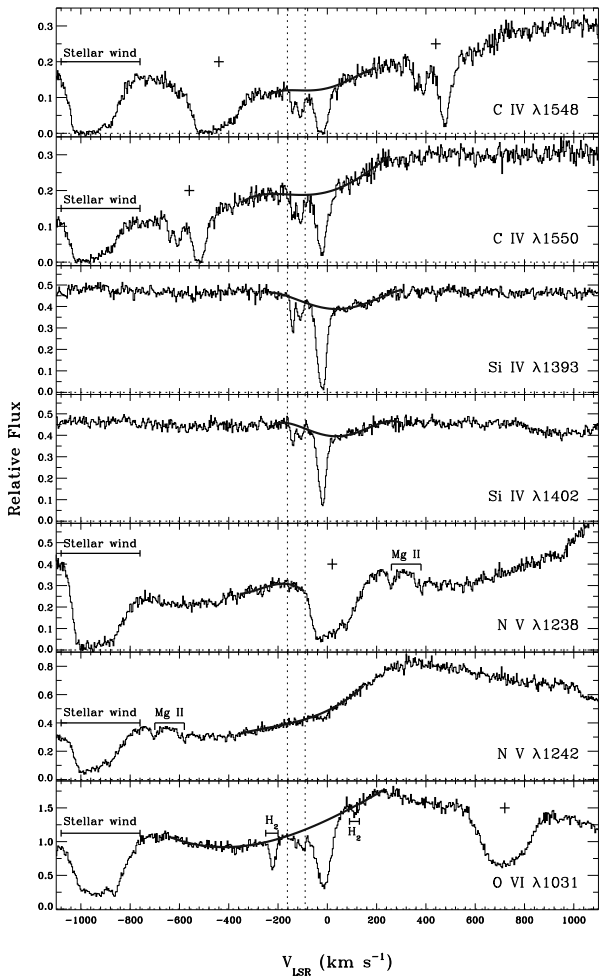}
\caption{Adopted continuum fits of the high ions are shown as the solid black 
curves.  The fits to the continua for the \ion{C}{4} and \ion{Si}{4} lines were adopted from 
the component fitting software.  The two vertical lines show the HVC region of 
interest.  The P-Cygni profile with a terminal velocity of $\sim 900$ \kms\ is 
clearly seen in the \civ, \nv, and \ovi\ lines.  The plus signs represent the 
profiles and/or stellar wind features of the other line of the doublet.  
We do not use \nv\ $\lambda1238$ in our analysis due to the difficulty in 
determining the continuum.  We show it here to demonstrate this difficulty 
and show the stellar wind profile.
\label{f3}}
\end{figure}

\zngi\ is a post-asymptotic giant branch star in the globular cluster M5 (NGC 5904); 
the properties of the star and cluster are summarized in Table~\ref{tab1}.  \zngi\ 
has a remarkably fast projected rotational velocity for a PAGB star 
($v \sin i = 170$ \kms).  One possible explanation may be that this star has 
been spun up by a merger with a binary companion.  This star has very little 
hydrogen in its atmosphere, with a helium abundance of 99\% by number 
(W. V. Dixon, unpublished).  Its photospheric carbon and nitrogen abundances 
are 10 times solar by mass, suggesting products of helium burning 
were mixed to the surface while the star was on the AGB 
(Dixon et al. 2004).  The sight line to this star shows high velocity absorption 
as first noted by Dixon et al. (2004).  Over the velocity range 
$-162$ $<$\vlsr$<-90$ \kms, absorption from the ions \HI, \cii, \civ, \nii, 
\oi, \ovi, \alii, \siii, \siiii, \siiv, \sthree, and \feii\ is seen in our 
\stis\ and \fuse\ observations (see Figures \ref{f1}, \ref{f2}, and \ref{f3}).  
The presence of strong \siiv, \civ, and \ovi\ absorption at high velocities 
($|v|\gtrsim100$ \kms) place these clouds in the category of highly-ionized 
HVCs recently discussed by Fox et al. (2005), Collins et al. (2004; 2005), and 
Ganguly et al. (2005).  High velocity absorption is detected in multiple 
components with centroid LSR velocities at roughly $-110$ \kms\ and $-140$ 
\kms\ ($-190$ \kms\ and $-160$ \kms\ relative to the photosphere of \zngi).  
Our aim is to determine the origin of this observed multiphase high-velocity 
absorption. 



\begin{deluxetable}{lcc}
\tabletypesize{\small}
\tablecolumns{3}
\tablewidth{0pt}
\tablecaption{Parameters of M5-ZNG~1
\label{tab1}}
\tablehead{
\colhead{Parameter} & 
\colhead{Value} &
\colhead{Reference} \tablenotemark{a}
}
\startdata
\cutinhead{Cluster Parameters}
Distance                & 7.5 kpc      & 3 \\
{[Fe/H]}                & $-1.27$      & 3 \\
$v_{{\rm {LSR}}}$       & $+65.7$ \kms\  & 3 \\
\cutinhead{Stellar Parameters}
Spectral type           & sdO           & 1 \\
$V$                     & 14.54          & 2 \\
\bv                   & $-0.32$        & 2 \\
$E$(\bv)                & 0.03          & 3 \\
$T_{\rm eff}$	&	$44300 \pm 300$ K & 4 \\
$\log g$	&	$4.3 \pm 0.1$ & 4 \\
$v \sin i$	&	$170 \pm 20$ km s$^{-1}$ & 4 \\
$v_{\rm LSR}$	&	$+51 \pm 3$ km s$^{-1}$ & 5 \\
$\log L/L_{\sun}$ &	$3.52 \pm 0.04$ & 4 \\
$M/M_{\sun}$ 	&	$0.69 \pm 0.17$ & 4 \\
C abundance\tablenotemark{b}	&	$2.58 \pm 0.20$\% & 4 \\
N abundance	&	$0.51 \pm 0.05$\% & 4 \\
O abundance     &       $0.37 \pm 0.32$\% & 4
\enddata
\tablenotetext{a}{References: (1) Napiwotzki \& Heber 1997; (2) Piotto 2002; 
(3) Harris 1996 (on-line version dated February 2003); (4) Dixon et al. 2004; 
(5) This work.}
\tablenotetext{b}{Abundances are quoted as mass fractions.}
\end{deluxetable}
%

The sight line to M5 \zngi\ probes a 7.5 kpc path toward the inner Galaxy.  It 
passes through Radio Loop I (centered at roughly $l=329\fdg0$, $b=17\fdg5$ with 
a diameter of $\sim 116^\circ$, Berkhuijsen et al. 1971).  The energy source for 
the Loop I superbubble is commonly thought to be stellar winds and/or supernovae 
in the Sco-Cen OB association located at a distance of $d\sim170$ pc from the 
sun (Park et al. 2007; Wolleben 2007 and references therein).\footnote{Loop I 
is typically associated with the Sco-Cen OB association; however, Bland-Hawthorn 
\& Cohen (2003) and Yao \& Wang (2007) argue that the North Polar Spur (NPS), 
which is usually thought to be a part of Loop I, is actually a limb brightened 
region associated with the outer walls of a bipolar Galactic nuclear wind.}  It 
is a priori possible that the high velocity absorption toward \zngi\ may be associated 
with material in Loop I.  

Savage \& Lehner (2006) analyzed the \ovi\ in the sight lines to 39 white dwarfs, 
of which 3 lie in the direction of Loop I at low latitudes with distances close 
to 200 pc.  No \ovi\ is observed in any of these sight lines at high velocities.  
Sembach et al. (1997) observed the sight line toward HD~119608 
($l=320\fdg4$, $b=43\fdg1$) that passes through Loop I.  In addition, they 
compiled archival \iue\ observations of $\sim20$ sight lines passing through 
Loop I.  Again, no high-ion absorption was detected at high velocities.  The 
sight line to 3C~273 passes through Loop I which shows very strong \civ\ and 
\ovi\ absorption with velocities between roughly $-100$ and $+100$ \kms\ 
(Sembach et al. 2001, 1997); a high velocity wing feature is also seen, 
reaching as far as $+240$ \kms\ (Sembach et al. 2001).  Given that the sight 
lines which probe \ovi\ in or near Loop I show no absorption at high negative 
velocities, it is unlikely that the high velocity gas toward \zngi\ 
represents Loop I material.  We note also that the [O/H] of the ISM within 
800 pc is observed to be nearly solar (Cartledge et al. 2004).  If the HVCs 
toward \zngi\ were to represent Loop I material, one would expect the 
metallicity to more closely match that of the local neighborhood.  The 
super-solar metallicity of the HVCs toward \zngi\ (see \S~\ref{sec:abundance}), 
combined with the absence of HVCs with \vlsr\ $\leq -100$ \kms, strongly suggest 
that they reside farther than the $\sim200$ pc distance of Loop I.

The Leiden-Argentine-Bonn (LAB) Survey (Kalberla et al. 2005) provides us with 
a neutral hydrogen column density of $N$(\HI)$=3.67 \times 10^{20}$ cm$^{-2}$ 
for \HI\ emission centered around \vlsr\ $\approx0$ \kms\ in the direction of M5.  
A perusal of the survey reveals no \HI\ emission at velocities 
$\left|\upsilon_{LSR}\right| \gtrsim 30$ \kms\ within $10^\circ$ of \zngi\ 
with a sensitivity of $N(\mbox{\HI})\approx3\times10^{18}$ cm$^{-2}$ for an 
HVC FWHM $\sim20$ \kms.\footnote{The brightness-temperature sensitivity of 
the LAB survey is 0.07--0.09 K with a sensitivity of 
$N(\mbox{\HI})\approx2\times10^{17}$ cm$^{-2}$ per 1.3 \kms\ resolution element.}  
The high velocity gas seen toward \zngi\ is not clearly associated with any 
known HVC complex (Wakker 2001).  The closest HVC complex is complex L 
($\Delta\theta\sim25^\circ-30^\circ$) which has a mean LSR velocity 
$\sim -115$ \kms.  Richter et al. (2005) find HVC absorption toward PKS~1448-232 
($l=335\fdg4$, $b=+31\fdg7$, $\Delta\theta\sim27^\circ$ from \zngi)
with LSR velocities near $-100$, $-130$, and $-150$ \kms\ that are similar 
to those observed in the HVCs toward \zngi.  Richter et al. attribute this 
absorption to complex L.  The compact HVC CHVC 018.3--47.1--147 
(de Heij et al. 2002) lies $\sim 10^\circ$ from \zngi\ with an LSR velocity of 
$-147$ \kms\ (this is cloud 57 in Wakker \& van Woerden 1991).

Zsarg\'{o} et al. (2003) analyzed \fuse\ spectra for 22 Galactic halo stars.  
They saw no clear evidence of high velocity \ovi\ in any of the sight lines.  
Their sample included \zngi; however, they used earlier \fuse\ 
data for this star with a S/N of about half of that of our data.  
\zngi\ is the only known Galactic sight line to show highly-ionized 
interstellar HVCs.

Sembach et al. (2003), in their survey of high velocity \ovi, analyzed 
highly-ionized high velocity gas along sight lines to 100 extragalactic 
objects and 2 halo stars.  Of the sight lines in their study, only Mrk~1383 
($l=349\fdg2$, $b=+55\fdg1$, $\Delta\theta=11\fdg8$ from \zngi) lies within 
$30^\circ$ of \zngi.  Mrk~1383 shows no blueshifted high velocity absorption 
(Keeney et al. 2006; Sembach et al. 2003).  Fox et al. (2006) analyzed the 
spectra of 66 extragalactic sight lines searching for highly-ionized HVCs.  
The sight line toward PG~1553+113 ($l=21\fdg9$, $b=+44.0$, $\Delta\theta=13\fdg1$) 
shows \ovi\ absorption at blueshifted velocities between $-170$ and $-100$ \kms\ 
(e.g., see Figure~6 of Fox et al. 2006, \S~8).  Fox et al. (2006) 
reported a $3\sigma$ upper limit for the \ovi\ absorption at negative LSR velocities 
($-200$ to $-100$ \kms).  The main difference between our analysis and theirs is the 
velocity range used to estimate the equivalent width and column density: Considering 
the velocity range \vlsr\ $=-200$ to $-100$ \kms, 
the $3\sigma$ upper limit on the equivalent width is 40 m\AA.  Between $-170$ and 
$-100$ \kms, we measure $W_\lambda = 39 \pm 11$ m\AA\ and $\log N(\mbox{\ovi}) = 
13.57^{+0.11}_{-0.13}$, corresponding to a 3.5$\sigma$ detection.

Given its evolutionary status, this star may be expected to have circumstellar 
material, perhaps even a (proto) planetary nebula (PN).  Dixon et al. (2004) have 
summarized the prior observations of \zngi\ and the search for a PN.  Bohlin et 
al. (1983) suggested that an apparent \ion{N}{4} $\lambda1478$ emission feature 
in the \iue\ spectrum of \zngi\ was a signature of a PN.  However, using an 
analysis of the complete archival \iue\ data, de Boer (1985) argued the apparent 
\ion{N}{4} feature was not real; he did note the P-Cygni profile of the 
\ion{N}{5}\ $\lambda\lambda$1238,1242 doublet, a sign of an outflowing wind 
from the star.  This wind shows a terminal velocity of $\sim900$ \kms\ (e.g., 
see Figure~\ref{f3} and Dixon et al. 2004).  Napiwotzki \& Heber (1997) set 
out to test the PN hypothesis by using \hst\ to obtain a WFPC2 \halpha\ image.  
The \halpha\ image revealed no evidence for extended emission around the star 
from a PN.  Any circumstellar \halpha\ is constrained to lie within $0.2''$ 
($7\times10^{-3}$ pc) from the star.  We will present strong evidence against 
the possibility that the HVCs in this direction are associated with 
circumstellar material in \S~\ref{sec:photoionizationmodel}.

At Galactic coordinates $l=3\fdg9$, $b=+47\fdg7$, a distance of $d=7.5$ kpc 
(Harris 1996), and a vertical distance above the Galactic plane of $z=+5.3$ 
kpc, \zngi\ lies about $3.5$ kpc from the rotation axis of the Galaxy.  At 
this location, a Galactic bipolar wind, for which evidence has been 
accumulating recently, may play a role in this HVC material.  Almy et al. 
(2000) analyzed \rosat\ X-ray observations toward Loop I and the Galactic 
center and were able to show that a large fraction (45\% $\pm$ 9\%) of the 
X-ray emission originates beyond $d\sim2$ kpc.  They suggested that the most 
probable source for this emission was the Galactic X-ray ``bulge.''  Yao \& 
Wang (2007) used a differential analysis of archival {\it Chandra} grating 
observations of the sight lines toward Mrk~421 ($l=179\fdg8$, $b=65\fdg0$) 
and 3C~273 ($l=290\fdg0$, $b=64\fdg4$) to obtain the net emission and 
absorption of the hot gas toward the Galactic center soft X-ray emission 
(seen toward 3C~273).  They showed that the X-ray emitting gas toward 3C~273 
originated beyond 200 pc and suggested the most likely source was a Galactic 
center outflow.  If this X-ray emission indeed originates from the Galactic 
center, then the Galactic latitude of 3C~273 ($b=65\fdg0$) 
suggests that an outflow can reach beyond the Galactic latitude of \zngi\ 
($b=47\fdg7$).  Based on the 408 MHz radio continuum and the \rosat\ all-sky 
soft X-ray data, Sofue (2000) calculated the expected size of an outflow.  His 
predictions for the outer boundary of an outflow encompass \zngi.  Furthermore, 
\zngi\ is located inside the outflow cones in the empirically-motivated models 
of Bland-Hawthorn \& Cohen (2003).  Keeney et al. (2006) have previously discussed 
a Galactic wind in the context of the highly-ionized HVCs toward Mrk~1383 and 
PKS~2005--489, attributing the HVCs to a Galactic center outflow.  The sight 
line toward \zngi\ may intercept material associated with feedback-driven flows 
in the central regions of the Galaxy, either from a nuclear wind or from Galactic 
fountain-type flows from the inner Galaxy.  We discuss this possibility in 
\S~\ref{sec:discussion}.


\section{Interstellar Absorption Line Measurements}
\label{sec:analysis}

We have measured column densities, equivalent widths, $b$-values, and 
signal-to-noise ratios for the ion profiles where significant absorption 
is seen in the HVC region.  Where the absorption is absent and not contaminated, 
$3\sigma$ upper limits for the column densities and equivalent widths were 
calculated.  We employed two principal methods in obtaining these values, the 
apparent optical depth (AOD) method as described by Savage \& Sembach (1991) and 
a component fitting method as described by Fitzpatrick \& Spitzer (1997), 
discussed in \S~\ref{sec:AOD} and \S~\ref{sec:compfit}, respectively.  We 
separately discuss our analysis of the \HI\ Lyman series absorption in 
\S~\ref{sec:HI}.


\subsection{AOD Measurements of the Metal Lines}
\label{sec:AOD}

Figures~\ref{f1} and \ref{f2} show the normalized intensity profiles for the 
low and high ions, respectively.  We normalized each absorption feature by 
fitting a low-order ($\leq5$) Legendre polynomial to the adjacent continuum.  
The continuum of \zngi\ for the most part was easily modeled, with the exception 
of the regions about \siiv\ and \civ.  The \civ\ $1548,1550$ \AA\ doublet lies 
in the presence of a stellar wind making the continuum placement more difficult.  
The \siiv\ $1393,1402$ \AA\ lines each lie in a stellar feature with the added 
complication that the \ion{Si}{4} $\lambda 1402$ line lies near the edge of a 
spectral order. The data from the spectral orders for the 1402 \AA\ line were 
coadded, however, there was a feature near $-85$ \kms\ which appeared in one 
order but not the other.  For these high ions, we adopt continua determined 
through a component fitting analysis, the details of which are described in 
\S~\ref{sec:compfit}.   Figure~\ref{f3} shows the adopted continua for the 
ions \civ\ $\lambda\lambda$1548, 1550, \siiv\ $\lambda\lambda$1393, 1402, 
\nv\ $\lambda\lambda$1238, 1242 and \ovi\ $\lambda$1031.  While continuum 
placement near the \ion{O}{6} $\lambda1031$ line can sometimes be problematic 
in stars, here the continuum is well determined.  We cannot use the \ovi\ 
$\lambda1037$ line because it is always contaminated by \cii, \cii$^*$, and \htwo.

The strong line of \ion{O}{6} at 1031 \AA\ can be contaminated by \htwo\ or HD 
lines from the Milky Way, and by \ion{Cl}{1} at 1031 \AA.  The \ion{Cl}{1} is 
at $-122$ \kms\ with respect to \ion{O}{6} which is in the velocity region of 
our HVCs.  We searched for absorption from \ion{Cl}{1} $\lambda\lambda$1004, 
1003, which have similar strengths to the 1031 transition, and found none.  Of 
the three molecular lines that can contaminate \ovi, HD 6--0 R(0) $\lambda1031$, 
(6--0) P(3) $\lambda1031$, and R(4) $\lambda1032$, are at $-4$, $-214$ and $+125$ 
\kms\ relative to \ovi, respectively, and are not in the velocity region of our HVCs.
 
Table~\ref{tab2} gives our measured properties of the high-velocity interstellar 
absorption lines toward \zngi, including the equivalent widths, apparent column 
densities, signal-to-noise, and $b$-values.  Table~\ref{tab3} gives the column 
densities and $b$-values for the two components seen in the high ions.  We 
measured equivalent widths of interstellar features following Sembach \& Savage 
(1992) including their treatment of the uncertainties.  We assumed statistical 
uncertainties for the \fuse\ data were dominated by the effects of fixed pattern 
noise, while the uncertainties for the \stis\ data were treated as statistical 
Poisson uncertainty.  In addition, we include an uncertainty in the placement of 
the continuum added in quadrature to the statistical uncertainty following Sembach 
\& Savage (1992).

\begin{deluxetable*}{lcccccccc}
\tabletypesize{\small}
\tablecolumns{9}
\tablewidth{0pt}
\tablecaption{Interstellar Absorption Lines Towards M5-ZNG 1
\label{tab2}}
\tablehead{
\colhead{Species} & 
\colhead{$\lambda$} & 
\colhead{$\log \lambda f$} &
\colhead{$W_\lambda$} & 
\colhead{$\log N_a$} &
\colhead{$\Delta v$\tablenotemark{a}} &
\colhead{$b_a$} &
\colhead{S/N\tablenotemark{b}} &
\colhead{Instrument} \\
\colhead{} & 
\colhead{[\AA]} & 
\colhead{} &
\colhead{[m\AA]} & 
\colhead{[cm$^{-2}$]} &
\colhead{[km s$^{-1}$]} &
\colhead{[km s$^{-1}$]} &
\colhead{} &
\colhead{}
}
\startdata
\ion{C}{2} & 1334.532 & 2.234  & $177\pm2$ & $>14.27$ & $-162,-90$ & $17.9\pm0.2$ & 31 & \stis \\
\ion{C}{2} & 1036.337 & 2.088  & $131\pm5$ & $>14.26$ & $-162,-90$ & $22.3\pm0.7$ & 16 & {\em\fuse}/LiF 1A \\
\ion{C}{2}$^\ast$ & 1335.708 & 2.186  & $<5$ & $<12.44$ & $-162,-90$ & \nodata & 42 & \stis\\
\ion{C}{4} & 1548.195 & 2.468  & $144\pm6$ & $13.70\pm0.04$ & $-162,-90$ & $24.4\pm1.0$ & 12 & \stis\\
\ion{C}{4} & 1550.770 & 2.167  & $92\pm6$ & $13.74\pm0.03$ & $-162,-90$ & $24.7\pm1.3$ & 14 & \stis\\
\ion{N}{1}\tablenotemark{c} & 1199.550 & 2.199  & $<12$ & $<12.82$ & $-162,-90$ & \nodata & 17 & \stis\\
\ion{N}{2} & 1083.994 & 2.079  & $102\pm12$ & $>14.07$ & $-162,-90$ & $23.9\pm2.5$ & 6 & {\em\fuse}/SiC 1A \\
\ion{N}{2}$^{**}$ & 1085.550 & 1.286  & $<29$ & $<14.19$ & $-162,-90$ & \nodata & 6 & {\em\fuse}/SiC 1A \\
\ion{N}{2}$^{**}$ & 1085.710 & 2.000  & $<25$ & $<13.41$ & $-162,-90$ & \nodata & 7 & {\em\fuse}/SiC 1A \\
\ion{N}{5} & 1242.804 & 1.985  & $<8$ & $<12.84$ & $-162,-90$ & \nodata & 27 & \stis\\
\ion{O}{1} & 1302.168 & 1.796  & $16\pm3$ & $13.38\pm0.08$ & $-162,-85$ & $20.8\pm5.0$ & 38 & \stis \\
\ion{O}{6} & 1031.926 & 2.136  & $33\pm3$ & $13.46\pm0.04$ & $-162,-90$ & $25.6\pm2.3$ & 23 & {\em\fuse}\tablenotemark{d} \\
\ion{Al}{2} & 1670.787 & 3.463 & $50\pm8$ & $12.11\pm0.07$ & $-162,-90$ & $20.4\pm3.6$ & 11 & \stis \\
\ion{Si}{2} & 1190.416 & 2.541  & $54\pm3$ & $>13.27$ & $-162,-110$ & $14.8\pm0.7$ & 18 & \stis \\
\ion{Si}{2} & 1193.290 & 2.842  & $86\pm4$ & $>13.23$ & $-162,-100$ & $15.7\pm0.7$ & 18 & \stis \\
\ion{Si}{2} & 1260.422 & 3.171  & $133\pm3$ & $>13.15$ & $-162,-90$ & $17.3\pm0.4$ & 37 & \stis \\
\ion{Si}{2} & 1304.370 & 2.052  & $36\pm2$ & $13.49\pm0.03$ & $-162,-90$ & $18.1\pm1.4$ & 40 & \stis \\
\ion{Si}{2} & 1526.707 & 2.307  & $63\pm3$ & $13.43\pm0.03$ & $-162,-90$ & $19.2\pm1.3$ & 29 & \stis\\
\ion{Si}{3} & 1206.500 & 3.293  & $197\pm3$ & $>13.30$ & $-162,-85$ & $22.0\pm0.3$ & 21 & \stis \\
\ion{Si}{4} & 1393.755 & 2.854  & $50\pm3$ & $12.81\pm0.02$ & $-162,-90$ & $23.3\pm1.0$ & 33 & \stis \\
\ion{Si}{4} & 1402.770 & 2.552  & $30\pm3$ & $12.86\pm0.04$ & $-162,-90$ & $21.8\pm1.8$ & 28 & \stis \\
\ion{S}{2} & 1253.811 & 1.135  & $<8$ & $<13.67$ & $-162,-90$ & \nodata & 28 & \stis \\
\ion{S}{2} & 1259.519 & 1.311  & $<6$ & $<13.39$ & $-162,-90$ & \nodata & 36 & \stis \\
\ion{S}{3} & 1190.208 & 1.421  & $12\pm3$\tablenotemark{e} & $13.67\pm0.10$\tablenotemark{e} & $-162,-105$ & $18.8\pm4.2$ & 18 & \stis \\
\ion{Fe}{2} & 1144.938 & 1.978  & $14\pm5$ & $13.19\pm0.13$ & $-162,-90$ & $24.7\pm8.2$ & 15 & {\em\fuse}/LiF 1B \\
\ion{Fe}{2} & 1144.938 & 1.978  & $13\pm5$ & $13.15\pm0.14$ & $-162,-90$ & \nodata & 16 & {\em\fuse}/LiF 2A \\
\ion{Fe}{2} & 1608.451 & 1.968  & $14\pm6$ & $13.07\pm0.02$ & $-162,-90$ & \nodata & 16 & \stis \\
\enddata
\tablecomments{Wavelengths and $f$-values are from Morton (2003).  Upper limits 
are $3\sigma$ estimates; a $b$-value of 20 \kms\ was assumed when no data were available.
}
\tablenotetext{a}{$\Delta \upsilon$ denotes the integration range.  Where the 
range differs from $-162$ to $-90$ \kms\ we have adjusted for
nearby contaminating absorption.}
\tablenotetext{b}{The quoted value is the signal to noise ratio per detector pixel.  
For \stis, the detector has a pixel size of $3.22$ \kms\ per pixel and \fuse, the 
detector has an output pixel size of $3.74$ \kms\ per pixel.}
\tablenotetext{c}{The other members of these multiplets were blended with a stellar 
or interstellar features and are not included in this table.}
\tablenotetext{d}{The \fuse\ data used for this measurement are the coaddition of 
data from the LiF 1A, Lif 1B, SiC 1A, and SiC 2B segments.}
\tablenotetext{e}{This line may be slightly contaminated by \ion{Si}{2}\ $\lambda 1190$.}
\end{deluxetable*}
\begin{deluxetable*}{lcccccc}
\tabletypesize{\small}
\tablecolumns{7}
\tablewidth{0pt}
\tablecaption{High Ion Component Integrations\tablenotemark{a}
\label{tab3}}
\tablehead{
\colhead{Species} & 
\colhead{$\lambda$} & 
\multicolumn{2}{c}{Component 1\tablenotemark{b}} &
\colhead{} & 
\multicolumn{2}{c}{Component 2\tablenotemark{c}} \\
\cline{3-4} \cline{6-7}
\colhead{} & 
\colhead{[\AA]} &
\colhead{$\log N_a$ [cm$^{-2}$]} & \colhead{$b_a$\tablenotemark{d} [km s$^{-1}$]} &
\colhead{} &  
\colhead{$\log N_a$ [cm$^{-2}$]} & \colhead{$b_a$\tablenotemark{d} [km s$^{-1}$]} 
}
\startdata
\ion{C}{4} & 1548.195  & $13.25\pm0.05$  & $11.0\pm1.1$ & & $13.50\pm0.02$  & $12.2\pm0.5$  \\
\ion{C}{4} & 1550.770  & $13.35\pm0.05$ & $12.2\pm1.3$ & & $13.53\pm0.04$ & $12.2\pm0.8$  \\
\ion{N}{5} & 1242.804  & $<12.73$ & \nodata\tablenotemark{e} & & $<12.75$ & \nodata  \\
\ion{O}{6} & 1031.926  & $12.92\pm0.08$ & $14.4\pm2.4$ & & $13.34\pm0.03$ & $15.6\pm0.8$ \\
\ion{Si}{4} & 1393.755 & $12.58\pm0.03$ & $7.8\pm1.1$ & & $12.44\pm0.04$ & $11.6\pm1.0$ \\
\ion{Si}{4} & 1402.770 & $12.65\pm0.05$ & $7.8\pm2.1$ & & $12.47\pm0.08$ & $9.9\pm2.0$  \\
\enddata
\tablenotetext{a}{These results were obtained by the AOD method.}
\tablenotetext{b}{The velocity range for component 1 is [$-162,-126$] \kms.}
\tablenotetext{c}{The velocity range for component 2 is [$-126,-90$] \kms.}
\tablenotetext{d}{The $b$-value quoted here is defined as 
$b_a = [2\int (v - \overline{v}_a)^2 N_a(v) dv / N_a]^{1/2}$ and 
integrated over the velocity range for each component.}
\tablenotetext{e}{$b$-values from \ion{O}{6} $\lambda$1031.926 LiF 1A 
were adopted in finding these upper limits.}
\end{deluxetable*}

The column densities quoted here are derived from the apparent optical depth 
$\tau_a(v)$.  The apparent optical depth is an instrumentally-blurred version 
of the true optical depth of an absorption feature and is given by
\begin{equation}
  \tau_a(v) = - \ln \left[ I(v)/I_c (v) \right],
  \label{eqn:tauv}
\end{equation}
where $I_c(v)$ is the estimated continuum intensity and $I(v)$ is the observed 
intensity of the line as a function of velocity.  The apparent column density 
per unit velocity, $N_a(v)$ [${\rm atoms \ cm^{-2} \ (km \ s^{-1})^{-1}}$], is 
related to the apparent optical depth by 
\begin{equation}
  N_a(v) = \frac{m_e c}{\pi e^2} \frac{\tau_a (v)}{f \lambda} = 3.768 \times 10^{14} \frac{\tau_a (v)}{f \lambda}, 
\end{equation}
where $\lambda$ is the wavelength in \AA, and $f$ is the atomic oscillator 
strength.  We adopt rest wavelengths and $f$-values from Morton (2003).  
Resolved saturated structure is not seen in any of the profiles, 
but if present, would be clearly identifiable.
Unresolved saturated structure can be identified by comparing the $N_a(v)$ 
profiles for different transitions of the same species; a smaller apparent 
column density in the stronger transition suggests saturation.  In regions 
of the profiles for which unresolved saturated structure is not significant, 
the integrated apparent column density, $N_a$, is equivalent to the true 
column density, $N$.  For cases where unresolved saturated structure becomes 
significant, the apparent column density is a lower limit to the true value.    
The integrated values of $v_a$, $b_a$, and log $N_a$ are obtained from 
$v_a = \int v N_a(v) dv/N_a$, $b_a = [2\int (v - \overline{v}_a)^2 N_a(v) dv / N_a]^{1/2}$, and $N_a = \int N_a(v) dv$, 
where the integration is performed over the absorption region noted in 
Table~\ref{tab2}.

According to Savage \& Sembach (1991), the AOD method is adequate for data with 
$b_{\rm line} \sim 0.25$--$0.50 b_{\rm inst}$, where $b_{\rm line}$ is the intrinsic 
$b$-value of the line and $b_{\rm instr}$ is the $b$-value of the instrument. 
Since $b \equiv$\,FWHM$/1.667$, for STIS E140M, $b_{\rm inst} \simeq 4$ \kms\ 
and for \fuse, $b_{\rm inst} \approx 12$ \kms. Since there is no tracer of cold 
gas in the HVCs along this line of sight, such as \cii*\ or \ion{C}{1}, it is 
very unlikely that there is any absorption line with $b\ll 1$ \kms.

When $\tau_a \ll 1$, unresolved saturation should not be problematic as long as 
$b$ is not much smaller than 1 \kms. For stronger lines, unresolved saturated 
structure can be identified by comparing the lines of the same species with 
different $f \lambda$. Following Savage \& Sembach (1991), the difference in 
$f\lambda$ must be a factor of 2 (or 0.3 dex) or more to be able to detect 
the effects of unresolved saturation.  If some moderate saturation exists, 
we can correct for it using the procedure described in Savage \& Sembach (1991).  
For various cases of blending and line broadening, they found a tight relation 
between the difference of the true column density and the apparent column density 
of the weak line against the difference of the strong line and weak line apparent 
column densities. The needed correction to the apparent column density of the 
weak line for a given difference between the strong- and weak-line apparent 
column densities are summarized in their Table~4.

For the \civ\ and \siiv\ doublets, the weak lines of the doublet give 
systematically larger $N_a$, suggesting that these lines suffer from small 
saturation (less than 0.02--0.05 dex), although as we argued above, the 
continua near these lines is complicated and errors in the continuum placement 
will have a greater effect on the weak lines. For \ovi, because the intrinsic 
broadening is large, this line is unlikely to be affected by saturation 
(see Wakker et al. 2003). Because the peak apparent optical depth for the 
\oi, \sthree, and \feii\ lines are $\la 0.16$, these lines are also unlikely 
to be saturated.  For \siii, 5 transitions are available: the strong lines at 
1190, 1193, and 1260 \AA\ show some saturation effects.  The apparent column 
density of \siii\ $\lambda$1304 (weakest line) is 0.06 dex larger than the 
one of \siii\ $\lambda$1526, although within 1$\sigma$ their apparent column 
densities overlap.  Nonetheless, this is an indication of weak saturation 
(the continuum placement near these lines is straightforward), and we 
therefore correct for it in our adopted column density by increasing the 
column density of \siii\ $\lambda$1304 by 0.06 dex (see Table 4 in Savage 
\& Sembach 1991). For \alii, we note that the peak apparent optical depth 
is smaller than for \siii\ $\lambda$1304 (0.5 compared to 0.7) and therefore 
the saturation correction is likely smaller than the error quoted in 
Table~\ref{tab2} (these two ions very likely probe the same gas given their 
similar ionization potentials).  Finally, for \cii, \siiii, \nii, only strong 
transitions are available and are likely all saturated.  We therefore only 
quote lower limits for these ions.

For lines which showed no significant absorption, we adopt $3\sigma$ upper 
limits for the equivalent width.  We follow Wakker et al. (1996) and calculate 
the limit on the equivalent width by
\begin{equation}
  \sigma(W)_{m\rm\AA} = 6.5\times 10^{-3} \frac{\lambda(\rm\AA)}{S/N}\sqrt{hb},
  \label{eqn:eqwidthsigma}
\end{equation}
where $\lambda$ is the rest wavelength, $S/N$ is the signal to noise ratio, 
$h$ is the velocity dispersion per pixel for the spectrograph (3.22 \kms\ 
for \stis\ E140M and 3.74 \kms\ for \fuse), and $b$ is the estimated $b$-value 
in \kms.  From this, the $3\sigma$ upper limits on the column density are 
calculated assuming that the lines fall on the linear part of the curve of growth.  

Our adopted total column densities (i.e. that include all the absorption 
between $-162$ and $-90$ \kms) are summarized in Table~\ref{tab4}.  Note 
that for \civ\ and \siiv, we adopt the results from the profile fitting 
described in the next section, while the measurement for $N($\HI$)$ is 
described in \S~\ref{sec:HI}.

\begin{deluxetable}{lcc}
\tabletypesize{\small}
\tablecolumns{3}
\tablewidth{0pt}
\tablecaption{Adopted Total Interstellar Column Densities 
\label{tab4}}
\tablehead{
\colhead{Species} &
\colhead{$\log N$\tablenotemark{a}} &
\colhead{Method}
}
\startdata
\HI         & $16.50\pm0.06$     & FIT,COG \\
\ion{C}{2}  & $>14.27$           & AOD \\
\ion{C}{2}$^*$  & $<12.44$      & $3\sigma$ \\
\ion{C}{4}  & $13.71\pm0.07$    & FIT \\
\ion{N}{1}  & $<12.82$          & $3\sigma$ \\
\ion{N}{2}  & $>14.07$          & AOD \\
\ion{N}{2}$^{**}$  & $<13.41$ & $3\sigma$ \\
\ion{N}{5}   & $<12.84$         & $3\sigma$ \\
\ion{O}{1}  & $13.38\pm0.08$    & AOD \\
\ion{O}{6}  & $13.46\pm0.04$    & AOD \\
\ion{Al}{2} & $12.11\pm0.07$    & AOD \\
\ion{Si}{2} & $13.55\pm0.03$    & AOD \\
\ion{Si}{3} & $>13.30$          & AOD \\
\ion{Si}{4} & $12.86\pm0.03$    & FIT \\
\ion{S}{2}  & $<13.39$          & $3\sigma$ \\
\ion{S}{3}  & $13.67\pm0.10$    & AOD \\
\ion{Fe}{2} & $13.07\pm0.02$    & AOD   \\
\enddata
\tablenotetext{a}{All upper limits are $3\sigma$ estimates and the velocity 
range is $-162$ to $-90$ \kms.  See \S~\ref{sec:analysis} for more details.}
\end{deluxetable}
%


\subsection{Component Fitting of the High Ions}
\label{sec:compfit}

The high ions \civ\ and \siiv\ are prominent in both HVC components toward \zngi.  
Since there may be overlap between the absorbing regions, we utilize the method of 
component fitting that allows us to separate the distinct velocity components.  
We employ software (described in Fitzpatrick \& Spitzer 1997) in which we 
construct a model for the absorption wherein each profile is composed of 
multiple Maxwellian ``clouds'' or components.  The best-fit values describing 
the gas are determined by comparing the model profiles convolved with an 
instrumental line-spread function (LSF) with the data.  The three parameters 
$N_i$, $b_i$, and $v_i$ for each component, $i$, are input as an initial 
guesses and subsequently varied to minimize $\chi^2$.  The fitting process 
enables us to find the best fit of the component structure using the data 
from one or more transitions of the same ionic species simultaneously.

\begin{figure}  
\epsscale{1.2}
\plotone{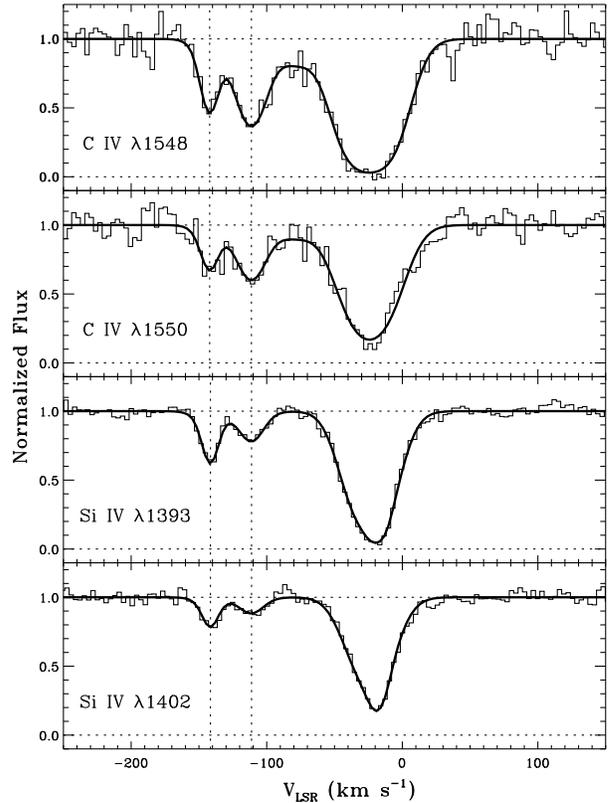}
\caption{The normalized profiles of the ions \ion{C}{4} and \ion{Si}{4} along 
with the best-fit component model (solid black line).  The centroid for component 
1 is $-142$ \kms\ and for component 2 is $-111$ \kms.  The vertical dotted lines 
represent the centroids derived for each ion. 
\label{f4}}
\end{figure}

We applied this component-fitting procedure to the \civ\ and \siiv\ doublets.  
In addition to fitting the profiles, we used the code to simultaneously 
determine the best fit continuum about each line (see Fitzpatrick \& 
Spitzer 1997).  The continuum for each of the four lines was modeled with 
a fourth-order Legendre polynomial.  The continua determined this way are 
shown as solid lines in Figure~\ref{f3}.  The \siiv\ and \civ\ ions were 
fit separately, i.e., we did not assume a common component structure for 
both ions a priori.  The profile fits are shown as solid lines in 
Figure~\ref{f4}.  Examination of the component models reveals a close 
agreement between the velocity centroids of \civ\ and \siiv\ in the HVC 
region.  We call the component at $\sim-142$ \kms\ {\it component 1}, and 
the component at $\sim -111$ \kms\ {\it component 2}.  In the low-velocity 
region of the \civ\ profiles, we allowed the software to determine, freely, 
components at \vlsr\ 
$\sim-79$ \kms\ and $\sim-24$ \kms\ to account for overlap of this low 
velocity material with the HVCs.  Similarly, for \siiv\ we included 
components at \vlsr\ $\sim-24$ \kms\ and $\sim-18$ \kms.

The results of the component fitting for \civ\ and \siiv\ are given in 
Table~\ref{tab5}.  The temperature, $T$, is determined by assuming only 
thermal broadening.  This gives $T\lesssim A(60.6)b^2$ where $A$ is the 
atomic weight of the ion, and $b$, the Doppler parameter in \kms, is 
obtained from the fit.  The temperatures of the gas are likely to be 
less than these values due to non-thermal motions.  These results are 
discussed in more detail in \S~\ref{sec:kinematics} in the context of 
kinematics and ionization.

\begin{deluxetable*}{lccccccccc}
\tabletypesize{\tiny}
\tablecolumns{10}
\tablewidth{0pt}
\tablecaption{Results of Component Fitting of High Ion Profiles
\label{tab5}}
\tablehead{
\colhead{Species} &
\multicolumn{4}{c}{Component 1\tablenotemark{a}} &
\colhead{} & 
\multicolumn{4}{c}{Component 2\tablenotemark{b}} \\
\cline{2-5} \cline{7-10}
\colhead{} &
\colhead{$V_{LSR}$} & \colhead{$\log N_a$} & \colhead{$b$} & \colhead{$T$\tablenotemark{c}} &
\colhead{} &  
\colhead{$V_{LSR}$} & \colhead{$\log N_a$} & \colhead{$b$} & \colhead{$T$\tablenotemark{c}} \\
\colhead{} &  
\colhead{[km s$^{-1}$]} & \colhead{[cm$^{-2}$]} & \colhead{[km s$^{-1}$]} & \colhead{[10$^5$ K]} &
\colhead{} &  
\colhead{[km s$^{-1}$]} & \colhead{[cm$^{-2}$]} & \colhead{[km s$^{-1}$]} & \colhead{[10$^5$ K]}
}
\startdata
\ion{C}{4}  & $-142.1\pm0.9$ & $13.25\pm0.06$ & $7.1\pm1.8$ & $<0.4\pm0.2$ &
            & $-111.3\pm1.4$ & $13.53\pm0.07$ & $12.8\pm2.7$ & $<1.2\pm0.5$ \\
\ion{Si}{4} & $-141.5\pm0.4$ & $12.63\pm0.03$ & $6.3\pm0.8$ & $<0.7\pm0.2$ &
            & $-111.2\pm0.9$ & $12.47\pm0.05$ & $10.6\pm1.8$ & $<1.9\pm0.6$  \\
\enddata
\tablenotetext{a}{The velocity range for component 1 is fixed as [$-162,-126$] \kms.}
\tablenotetext{b}{The velocity range for component 2 is fixed as [$-126,-90$] \kms.}
\tablenotetext{c}{$T$ is determined by using $T\lesssim A(60.6)b^2$ where $A$ is 
the atomic weight and [$b$] = \kms. 
Using both species, we solve for the temperature and the turbulent velocity for 
each component.  We find for component 1:
$T\sim1.4\times10^4$ K, $v_{nt}\sim5.6$ \kms; for component 2: $T\sim6.5\times10^4$ 
K, $v_{nt}\sim8.6$ \kms.}
\end{deluxetable*}
%

Since there is a good agreement in the velocity structure between the \civ\ 
and \siiv\ ions in the HVC region, we looked for the same velocity structure 
in \ovi\ (note that \nv\ is not detected).  Although we were able to fit the 
HVC region of \ovi\ $\lambda$ 1031 with one component (\vlsr\ $=-101.2\pm2.8$ 
\kms, $b=35.3\pm11.5$ \kms, $\log{N}=13.62\pm0.04$ cm$^{-2}$), we were unable 
to satisfactorily obtain a two-component fit likely due to both the cruder 
\fuse\ resolution and an intrinsically different velocity distribution of 
\ovi.  We discuss the velocity structure of \ovi\ along with the other ions 
in \S~\ref{sec:kinematics}.


\subsection{\HI\ Column Density Measurements}
\label{sec:HI}

\begin{figure}
\epsscale{1.0} 
\plotone{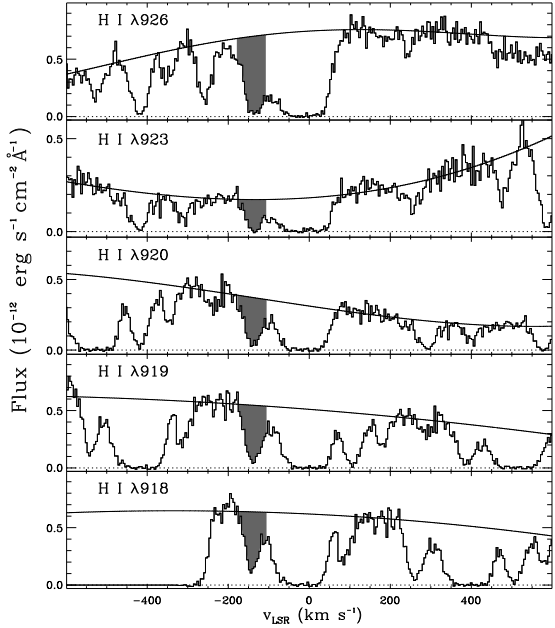}
\caption{{\fuse}\ SiC\,2A spectra of the \HI\ transitions used to estimate the HVC 
\HI\ column density 
(filled absorption at $\sim$$-140$ \kms). The solid thick lines shows our 
model for the continuum. 
The data are binned by 4.2 \kms\ per pixels, the {\fuse}\ resolution at 
these wavelengths is $\sim$25 \kms\ (FWHM). 
 \label{f5}}
\end{figure}

To estimate the \HI\ column density of the HVC, we used the Lyman series from 
926 down to 918 \AA,
where the HVC absorption is separated from the stronger \HI\ absorption at 
higher velocities. 
In Figure~\ref{f5}, we show the \HI\ profiles of the transitions 
considered in our measurements from the SiC\,2A detector. Higher wavelength 
transitions were not used
because all the components were blended (i.e., the Galactic component is so strong
that it covers the weaker absorption at lower velocities). The continuum shown in 
Figure~\ref{f5} for each transitions was estimated over a large range of velocities 
in order to reproduce the overall stellar continuum. The same 
procedure for the continuum placement was employed for the SiC\,1B detector segment. 

\begin{figure}
\epsscale{1.1} 
\plotone{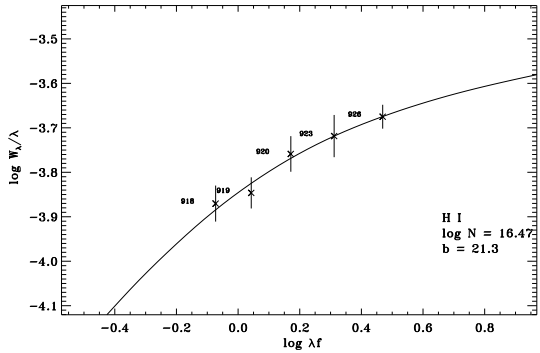}
\caption{Best fit single-component COG for the HVC absorption produced (\HI\ 
with $b = 21.3\,^{+2.7}_{-2.4}$ \kms\ and $\log N($\HI$)=16.47 \,^{+0.11}_{-0.08}$).  
The equivalent widths represent the averages of measurements from the {\fuse}\
SiC\,1B and SiC\,2A. 
\label{f6}}
\end{figure}

To estimate the column densities, a curve-of-growth (COG) method
and a profile fitting method were used. The COG method used the minimization
of the $\chi^2$ error derivation approach summarized by Savage et al. (1990).
In Table~\ref{tab6}, we summarize the averaged equivalent width measurements 
estimated in the SiC\,2A and SiC\,1B profiles. The 1$\sigma$ errors 
include continuum and statistical errors. The darkened
part of the spectra in Figure~\ref{f5} shows the typical integration range
where the equivalent width was measured. The result from 
the single-component COG is shown in Figure~\ref{f6}, where $b = 21.3\,^{+2.7}_{-2.4}$ 
\kms\ 
and $\log N($\HI$)=16.47 \,^{+0.11}_{-0.08}$. 

\begin{deluxetable}{lcc}
\tabcolsep=3pt
\tablecolumns{3}
\tablewidth{0pt} 
\tabletypesize{\footnotesize}
\tablecaption{\HI\ Equivalent Widths for the HVC \label{tab6}} 
\tablehead{\colhead{$\lambda$}    &  \colhead{$f$}  &  \colhead{$W_{\lambda}$} \\ 
\colhead{[\AA]}  &\colhead{}&\colhead{[m\AA]}}
\startdata
926.2257   & $ 3.18\times 10^{-3} $ & $ 195.8 \pm  12.5  $   \\
923.1504   & $ 2.22\times 10^{-3} $ & $ 176.5  \pm 20.4  $   \\
920.9631   & $ 1.61\times 10^{-3} $ & $ 160.5  \pm 15.5  $   \\
919.3514   & $ 1.20\times 10^{-3} $ & $ 131.0  \pm 11.0  $   \\
918.1294   & $ 9.21\times 10^{-4} $ & $ 123.8  \pm 12.1  $   \\
\enddata
\tablecomments{The equivalent widths represent the average of the SiC\,2A and SiC\,1B
measurements. We adopt the $f$-values from Morton (2003).}		
\end{deluxetable}

\begin{figure*}
\epsscale{1.0} 
\plotone{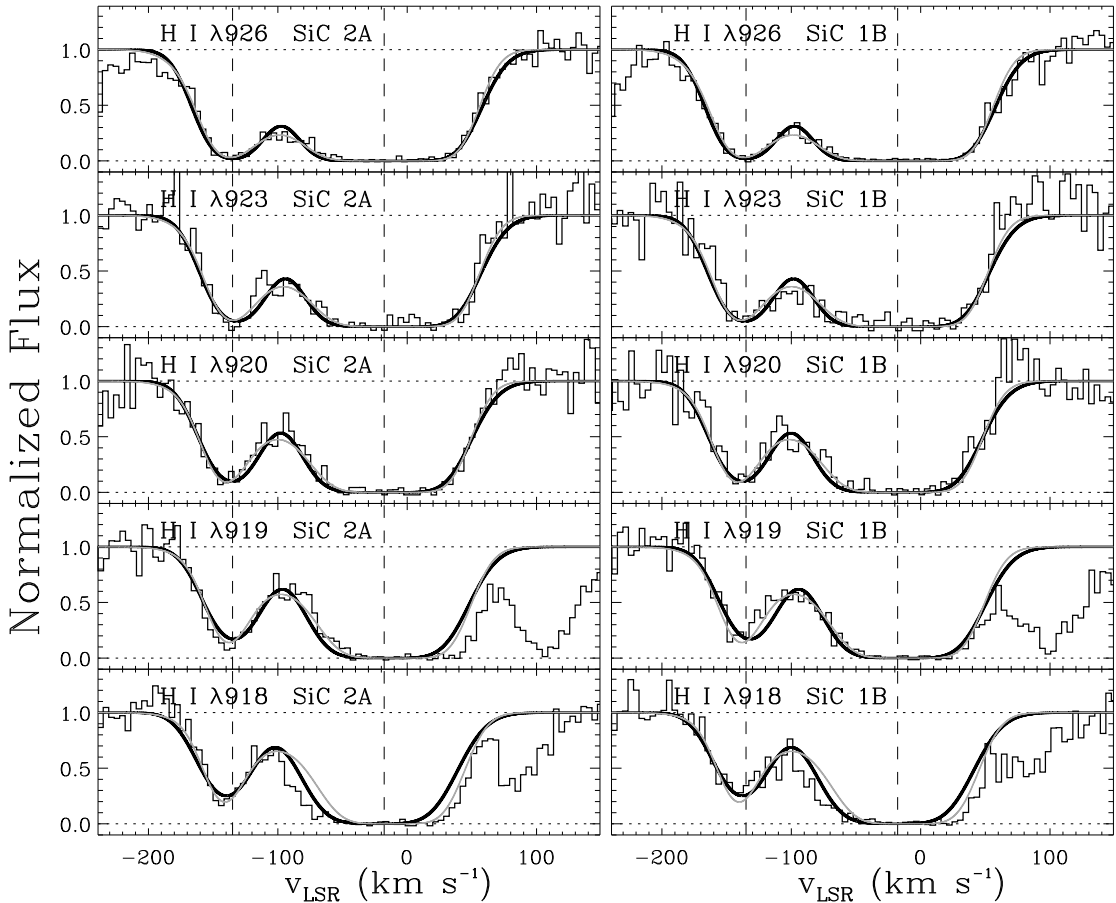}
\caption{Normalized \HI\ profiles against the LSR velocities. The left-hand 
side shows
data from SiC\,2A, the right-hand side shows data from SiC\,1B. The black 
solid line shows the fit
to the \HI\ transitions where the velocity centroid of each component is 
depicted by the dashed line. 
The grey dotted lines show an example of a fit with 4 components, where 
the velocities are 
$-140$, $-125$, $-59$, $-7.9$ \kms\ as suggested by the metal lines.  
In the two-component fit, the total 
\HI\ column density
for the HVC is 16.48 dex, while in the four-component fit, it is 16.58 dex. 
Note that the excess absorption in the \HI\ $\lambda$$\lambda$918, 919 
lines is due to contamination 
from other absorption lines.
\label{f7}}
\end{figure*}

For the profile fitting, we assume a Gaussian instrumental line spread function with a 
${\rm FWHM} = 25$ \kms. We simultaneously fit all the \HI\ transitions 
summarized in Table~\ref{tab6} for both segments, SiC\,2A and SiC\,1B. In Figure~\ref{f7}, 
we show an example of this process, where we fitted the absorption with one HVC component
and one low velocity component. The result of the fit gives for the HVC component 
$v = -135.0 \pm 0.8$ \kms, 
$b = 21.5 \pm 0.5$ \kms, and $\log N = 16.48 \pm 0.02$, which is consistent with the COG 
result although with much smaller errors. However, the absorption in both the local 
and HVC shows multiple components. Considering the \oi\ $\lambda$1302 profile (which is 
the best 
proxy for the \HI\ profiles), the HVC has at least two components, at $-140$ \kms\ and 
$-125$ \kms. 
Considering the \sii\ profile, the local absorption reveals at least two components. 
Importantly, the singly-ionized species systematically show intermediate-velocity 
absorption at $-59$ \kms, and from our profile fitting trials (see below), its strength 
and broadening can affect the total column density of the HVC. 
In the \none\ absorption profile, this component is not observed. In the \oi\ $\lambda$1302
profile, the $-59$ \kms\ component is blended with \pii\ $\lambda$1301. Unfortunately, 
\pii\ $\lambda$1152 has an apparent optical depth near unity; thus the 0.3 dex larger 
column density measured in  \pii\ $\lambda$1301 relative to \pii\ $\lambda$1152 is 
consistent
with some saturation in the latter transition or with contamination of the \pii\ 
$\lambda$1301 
by intermediate velocity absorption from \oi\ $\lambda$1302.  Therefore, it is 
not clear how much 
\oi\ absorption there is at $-59$ \kms, if any. The \oi\ profiles in the {\fuse}\ 
bandpass provide no additional 
information.  There is likely some \HI\ absorption at $-59$ \kms, but since the 
singly-ionized species trace both neutral and ionized gas, 
it is not clear how strong and broad the \HI\ absorption is in this region. 

To test the robustness of the previous fit with 2 components, we modeled the \HI\ 
lines using several combinations of parameters, including fits with 5 components
centered at $-140$, $-125$, $-59$, $-6$, $+12$ \kms\ and  4 components 
(removing the $+12$ \kms\ component), 
where all the parameters were allowed to vary (except the velocity at $-59$ \kms) or where
some parameters were fixed (for example, $b$ and $v$ in the $-59$ \kms\ component). 
We show in Figure~\ref{f7} an example of a four-component fit where the all the 
parameters were allowed
to vary except for the velocity centroids of the HVC and intermediate-velocity cloud. 
The result of our robustness trial is that the total \HI\ column density in the HVC ranges 
from 16.44 to 16.60 dex depending on the assumptions. The reduced-$\chi^2$  
is within 10\% in all our profile fitting trials. Our fitting results
give a mean \HI\ column density of  $\log N($\HI$)=16.52 \pm 0.08$. Since the 
COG and fit methods 
explore different $\chi^2$ parameters, we adopt a weighted average of the 
COG and profile fitting
results. Our adopted weighted-average (total) \HI\ column density of the HVC is
$\log N($\HI$)=16.50 \pm 0.06$.

We note that the single-component fit to the \HI\ provides a firm upper limit 
to the temperature of component 1: $T\leq (2.80\pm0.13)\times 10^4$ K.  Because 
this single component is known to be a combination of multiple absorbing regions, 
the non-thermal broadening is significant.  The \HI-bearing gas making up 
component 1 is thus quite cool.


\section{Physical Conditions and Chemical Abundances in the HVC}
\label{sec:physicalconditionsabundances}


\subsection{Electron Density}
\label{sec:electrondensity}

We make use of the \ion{C}{2} $\lambda$1334 and \ion{C}{2}$^*$
$\lambda$1335 lines to calculate an upper limit to the electron
density.  The \ion{C}{2}$^*$ line arises out of the $^2P_{3/2}$ upper fine
structure level, which has an energy $E_{12}\sim8\times10^{-5}$ eV
above the $^2P_{1/2}$ ground state, out of which the \ion{C}{2} transition
arises.  The excitation in regions of at least moderate ionization
(see \S~\ref{sec:ionizationfraction}) is dominated by electrons, while
the deexcitation should proceed principally via spontaneous radiative
decay.  Following Spitzer's (1978) discussion of the excitation
balance, the electron density is related to the column density ratio
$N(\mbox{\cii}^*)/N(\mbox{\cii})$ by (Lehner et al. 2004):
\begin{equation}
   n_e = 0.53 \, \frac{T^{1/2}}{\Omega_{12}(T)}e^{(E_{12}/kT)}
                    \frac{N(\mbox{\cii}^*)}{N(\mbox{\cii})}
  \label{eqn:e_density}
\end{equation}
where $\Omega_{12}(T)$ is the temperature-dependent collision strength
from Blum \& Pradhan (1992) and Keenan et al. (1986).  This yields an
upper limit to the electron density because the column of \ion{C}{2}
$\lambda$1334 is a lower limit due to saturation and \ion{C}{2}$^*$
$\lambda$1335 is an upper limit.  Using the limit to \ion{C}{2}$^*$
for the combined HVCs toward \zngi\ we find $n_e \lesssim (0.36$
cm$^{-3})T_4^{1/2}$ ($3\sigma$) for $T_4\equiv T/(10^4\rm{K})\sim1$ to
5 within a few percent.  The density limit rises to $n_e \la1.5$
cm$^{-3}$ for $T_4=10$, but the collision strengths for $T_4\ga5$ are
extrapolated from the lower temperature calculations and thus are
uncertain.  At such temperatures, the ionization fraction of \cii\
should be small, as well.

\ion{Si}{2} may be used as a proxy for \cii.  If one assumes solar
relative abundances, the implied column density of \cii\ is $\log
N(\mbox{\ion {C}{2}}) \approx 14.43$.  Depletion is unlikely to affect
this estimate much, as \ion{Si}{2} and \cii\ likely have similar
depletion characteristics in clouds with ``halo cloud'' (Sembach \&
Savage 1996) abundances.  Differing ionization levels for C and Si
could affect this determination.  Adopting $\log N(\mbox{\ion {C}{2}})
= 14.43$, we find $n_e \la (0.25 \ {\rm cm}^{-3})\,T_4^{1/2}$
($3\sigma$) for the high velocity gas.

These estimates assume \cii$^*$ is distributed like \cii, where
$\ga$ 2/3 of the column is associated with component 1.  If one
individually assesses the densities for components 1 and 2, the limits
($3\sigma$) are $n_e \lesssim (0.55$ cm$^{-3})T_4^{1/2}$ and
$\lesssim (0.95$ cm$^{-3})T_4^{1/2}$ for components 1 and 2,
respectively.

\subsection{Gas-Phase Abundance}
\label{sec:abundance}
Since \ion{O}{1}\ and \ion{H}{1}\ have nearly identical ionization potentials
and are strongly coupled through charge exchange reactions (Field \& Steigman
1971), $N$(\ion{O}{1})/$N$(\ion{H}{1}) can be used as a reliable metallicity
indicator.  The ionization corrections relating \ion{O}{1}/\ion{H}{1} to O/H
are extremely small.

We have measured a total \oi\ column density of $\log N$(\ion{O}{1})
$=13.38 \pm 0.08$ and a total \HI\ column density of $\log
N$(\ion{H}{1}) $=16.50 \pm 0.06$.  Thus the abundance of oxygen is
$\log{\rm{O/H}}=-3.12\pm0.10$.  Assuming the solar abundance of oxygen
is $\log{\rm{O/H}}=-3.34\pm0.05$ (Asplund et al. 2005), the gas phase
abundance of the high velocity gas is then [O/H] $=+0.22 \pm 0.10$,
where [X/Y] $= \log[N_X/N_Y] - \log\{X/Y\}_\odot$.\footnote{The older
  solar system abundances of Grevesse \& Noels (1992) give [O/H]
  $=+0.01\pm0.10$.}  Thus, the neutral gas in the HVCs toward \zngi\
has a super-solar oxygen abundance, the highest measured for any HVC.
We assume this abundance is representative of the highly-ionized HVC
gas as well.

The measured HVC abundance is high compared with the current best
solar abundance, the interstellar gas-phase abundance of oxygen in the
solar neighborhood ($\log{\langle \rm{O/H} \rangle}=-3.41\pm0.01$,
which is likely affected by modest depletion; Cartledge et al. 2004),
the average abundance of young F and G dwarfs in the solar
neighborhood ($\log{\langle \rm{O/H} \rangle}=-3.35\pm0.15$; Sofia \&
Meyer 2001), and the abundances of \ion{H}{2} regions in the solar
neighborhood ($\log{\langle \rm{O/H} \rangle}=-3.58$ to $-3.33$;
Rudolph et al. 2006 and references therein).  While determinations of
the solar oxygen abundance have varied significantly recently, the
gas-phase oxygen abundance in the HVCs toward \zngi\ is larger than, or
at least consistent, with all of the determinations.  These HVCs have
significantly higher metallicities than all other known HVCs, and 
the gas found locally in the Galactic disk.  Depletion effects are
unlikely to be significant in this determination; if depletion into
the solid phase is important, the total gas+dust phase abundance of
the HVCs would be even larger.


\subsection{Ionization Fraction}
\label{sec:ionizationfraction}

We can estimate the fraction of the gas that is ionized by comparing
the \HI\ column with an estimate of the total column density of
hydrogen ($N({\rm H}) \equiv N(\mbox{\HI}) + N(\mbox{\ion{H}{2}})$,
where we assume H$_2$ is negligible).  We estimate the total H column
from the total column of Si: $N({\rm H}) \approx N({\rm Si}) \, {\rm
(Si/H)}^{-1}$, where $N({\rm Si}) \ga N(\mbox{\siii}) +
N(\mbox{\ion{Si}{3}}) + N(\mbox{\ion{Si}{4}})$.  The inequality arises
because \ion{Si}{3} is saturated, and we do not have observations of
higher stages of ionization (though these should make little
contribution).  We adopt $\log ({\rm Si/H}) = \log ({\rm Si/H})_\sun +
0.22 = -4.27$, which assumes solar relative Si/O and a base Si
abundance from Asplund et al. (2005).  In regions with significant 
dust, the gas-phase Si/O may be subsolar.  We assume the conditions 
in this gas are such that depletion of Si/O is not significant.  
The total H column derived in
this way using the column densities in Table~\ref{tab4} is $\log
N({\rm H}) \ga 18.03 $, giving $\log { N(\mbox{\HI})/N({\rm H})} \la
-1.48$.  The total H column is $\gtrsim30$ times that of the neutral
column, and the ionization fraction is then $x(\rm{H}^+)\equiv
N(\rm{H}^+)/N(\rm{H}) \ga 0.97$.  This large ionization fraction
implies $N_{\rm{H}} \approx N_{\rm{H^+}} \approx N_{e}$, and justifies
our assumption in \S~\ref{sec:photoionizationmodel} that $n_{\rm
  H}\approx n_e$.

This represents the value for the HVCs integrated over components 1
and 2.  Very similar values are derived if the components are taken
individually.  Assuming the velocity structure of \HI\ follows that of
\oi, since their ionization fractions are locked together by a strong
charge exchange reaction, we use the \oi\ columns integrated over the
velocity ranges of components 1 and 2 to estimate \HI\ in each
component assuming a constant $\log ({\rm O/H}) = -3.12$ 
(\S~\ref{sec:abundance}).  This approach yields $\log {
  N(\mbox{\HI})/N({\rm H})} \la -1.44$ and $-1.56$ for components 1
and 2, respectively, implying ionization fractions $x(\rm{H}^+) \ga
0.96$ and $\ga0.97$.

\subsection{Path Length Through the HVCs}
\label{sec:pathlength}

The results of the previous subsections can be used to estimate the
path length through the HVCs.  The path length through a cloud is
$\Delta l \approx N({\rm H}) n_{\rm H}^{-1}$ (this assumes constant 
$n_{\rm H}$, though we discuss clumpy media below).  Because we do not
measure the total hydrogen column, $N({\rm H})$, we use Si or C as a
proxy for H (as in \S~\ref{sec:ionizationfraction}): $\Delta l_{\rm
  Si} \approx N({\rm Si}) \, {\rm (Si/H)}^{-1} \, n_{\rm H}^{-1}$.
Using the electron density limits from \S~\ref{sec:electrondensity}
for $n_{\rm H}$, we derive $\Delta l_{\rm Si} > (1.1 \ {\rm
  pc})\,T_4^{-1/2}$ ($3\sigma$) for the integrated HVCs.  For
temperatures consistent with the temperature limits on component 1
(see Table \ref{tab5}), this implies $\Delta l_{\rm Si} > 0.6$ pc.
Similar values can be derived using the columns of \ion{C}{2} and
\ion{C}{4} ($\Delta l_{\rm C} > (0.5 \ {\rm pc})\,T_4^{-1/2}$).  These
estimates apply to both components of the HVC, following the
discussion of \S~\ref{sec:electrondensity}, and do not depend on
ionization assumptions.

Taking components 1 and 2 individually, and using the electron density
limits derived for each in \S~\ref{sec:electrondensity}, we find
limits ($3\sigma$) of $\Delta l_{\rm Si} > (0.25 \ {\rm
  pc})\,T_4^{-1/2}$ and $ > (0.15 \ {\rm pc})\,T_4^{-1/2}$ for
components 1 and 2.

Clumping of the gas into regions of different densities cannot lower
this value, as the majority of the gas seen in these ions must be at
densities consistent with the limits from \cii$^*$.  If a large
fraction of the column of gas were in high density clumps, \cii$^*$
absorption would be present.  These path lengths are inconsistent with
typical sizes of circumstellar matter (e.g., discussion in Sahai et
al. 2007).  We discuss this further in Section 
\ref{sec:photoionizationmodel}.


\section{Velocity Structure, Kinematics, and Ionization Structure}
\label{sec:kinematics}


\subsection{Profile Comparisons}
\label{sec:profilecomparisons}

\begin{figure}
\epsscale{1.2}
\plotone{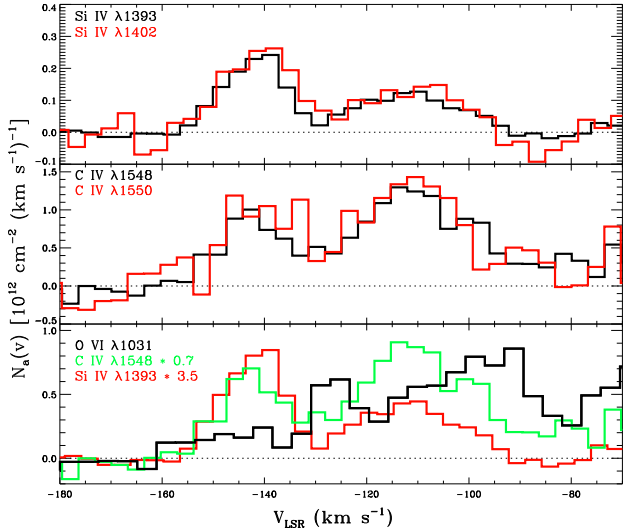}
\caption{A comparison between the apparent column densities of the high ions 
in the ZNG1 sight line.  The top two panels show both lines of the \ion{Si}{4} 
and \ion{C}{4} doublets.  The good agreement between the profiles of each 
member of the doublet implies little saturation.  The bottom panel shows 
\ion{O}{6} with scaled versions of \ion{Si}{4} and \ion{C}{4}.
\label{f8}}
\end{figure}

Comparing the apparent column density profiles of the high and low ions provides a 
visual means by which we can examine the kinematic structure of these species.  
Figure~\ref{f8} shows the apparent column density vs. \vlsr\ for the high ions 
\siiv, \civ, and \ovi.  Two velocity components are clearly observed in \ion{Si}{4} 
and \ion{C}{4} with LSR centroid velocities of roughly $-140$ and $-110$ \kms.  
The centroid velocities of the components coincide well between the two ions 
(see Table~\ref{tab5}) indicating that these ions reside in the same gas.

\begin{figure}  
\epsscale{1.2}
\plotone{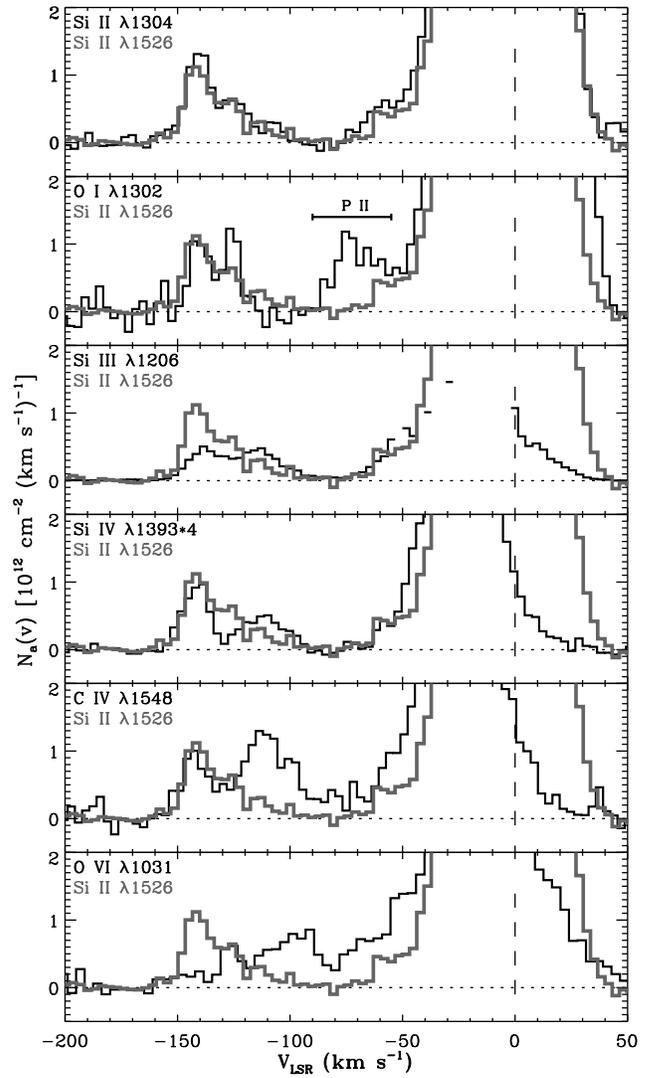}
\caption{Apparent column densities of various ions compared with \siii\ 
$\lambda1526$.  The $N_a(v)$ profile for \siiv\ $\lambda1393$ is multiplied 
by a factor of four.  Components 1 and 2 have velocity ranges of $-162$ to $-126$ 
\kms\ and $-126$ to $-90$ \kms, respectively.  The centroid velocities are 
$-142$ and $-111$ \kms.  The ions \siii\ and \oi\ appear to show a third component 
with a centroid velocity of $\sim-125$ \kms.
\label{f9}}
\end{figure}

\begin{figure}  
\epsscale{1.0}
\plotone{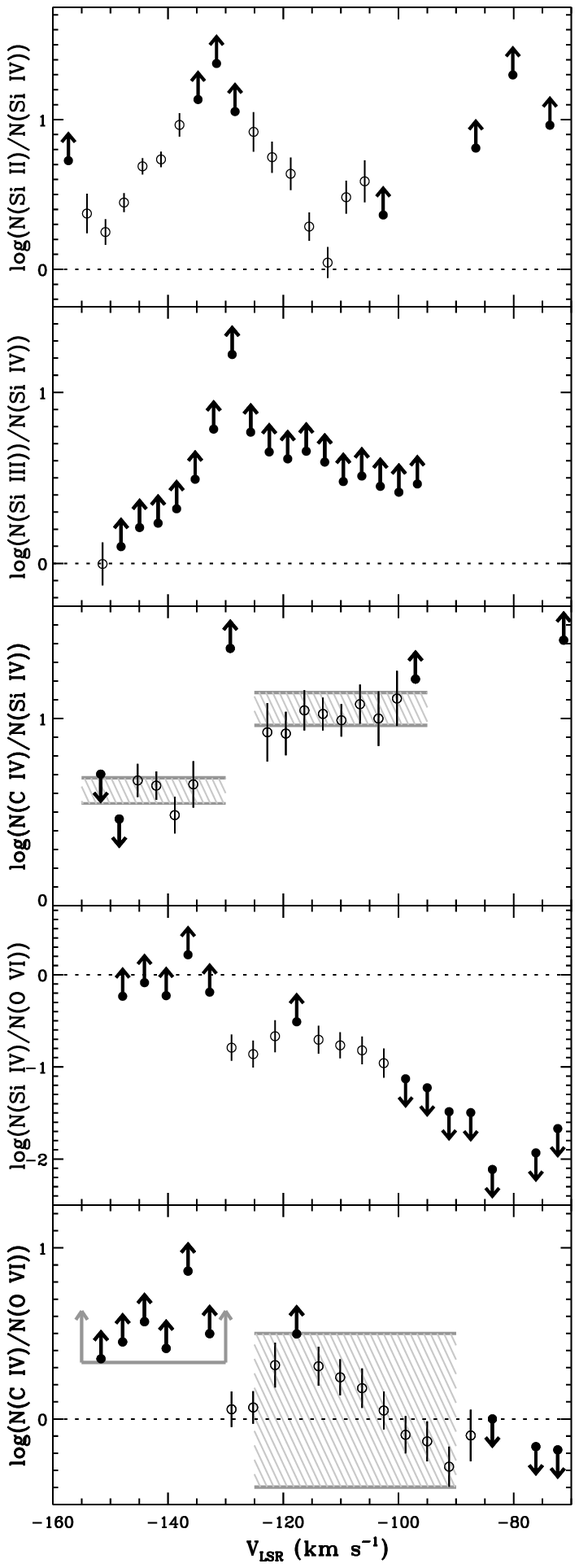}
\caption{Logarithmic apparent column density ratios as a function of \vlsr.  The ions plotted 
are \ion{Si}{2}\
$\lambda1304$, \ion{Si}{3}\ $\lambda1206$, \ion{Si}{4}\ $\lambda1393$, \ion{C}{4}\
$\lambda1548$, and \ion{O}{6}\ $\lambda1031$.  The upward (downward) arrows 
represent lower (upper) limits.  The hatched regions in the \civ/\siiv\ ratio 
represent the values derived from component fitting (see Table~\ref{tab7}).  
The hatched region in the \civ/\ovi\ ratio represents the maximum and minimum 
values for this ratio, although clearly a trend is seen as the ratio decreases 
toward higher velocities.  The light gray arrows in the \civ/\ovi\ ratio 
represents the lower limit for component 1 (see Table~\ref{tab7}).  
Components 1 and 2 have velocity ranges of $-162$ to $-126$ 
\kms\ and $-126$ to $-90$ \kms, respectively.
\label{f10}}
\end{figure}

The \ion{C}{4}/\ion{Si}{4} ratio changes significantly from component 1 to 
component 2 (Figure~\ref{f8}).  \ion{Si}{4} is more prominent in component 1 
than component 2 (log $N_a = 12.63\pm0.03$, log $N_a = 12.47\pm0.05$, 
respectively), but the opposite is true for \civ\ (log $N_a = 13.25\pm0.06$, 
log $N_a = 13.53\pm0.07$, respectively).  Since \civ\ has a greater ionization 
potential than \siiv, this suggests component 2 has a higher degree of ionization, 
perhaps implying higher temperatures which would be consistent with the results of 
component fitting (Table~\ref{tab5}).  This is also supported by the increasing 
\ovi\ and decreasing low ion contributions.  From Figures \ref{f9} and \ref{f10} 
we see that more \siiii\ resides in component 2 than does \siiv; yet there is 
more \civ\ than \siiv\ in that component.  The ionization potentials for \siiii, 
\siiv, and \civ\ are 16.3, 33.5, and 47.9 eV, respectively, which suggests that 
there are at least two processes involved in ionizing that component.  The 
bottom panel of Figure~\ref{f8} shows the apparent column densities vs. \vlsr\ 
for the strong lines of \civ, \siiv, and \ovi.  The velocity distribution of \ovi\ 
deviates from the common component structure seen in \civ\ and \siiv, with an 
increasing column from \vlsr\ $\sim-120$ to $-90$ \kms.  As the columns of 
\civ\ and \siiv\ start to decrease at $\sim -110$ \kms, $N(\mbox{\ovi})$ 
continues to {\it increase} up to $\sim -90$ \kms, perhaps suggestive of an 
increase in temperature with velocity leading to higher ionization states. 


Figure~\ref{f9} shows a comparison of the apparent column density profiles of 
\siii\ $\lambda$1526 with the ions \siii\ $\lambda$1304, \oi\ $\lambda$1302, 
\siiii\ $\lambda$1206, \siiv\ $\lambda$1393, \civ\ $\lambda$1550, and \ovi\ 
$\lambda$1031.  We note that the low ions are mostly confined to \vlsr\ 
$\lesssim-125$ \kms.  The low ions reveal a third component with a centroid 
velocity of roughly $-127$ \kms\ which is not obvious in the profiles of 
\civ\ and \siiv.  This third component lies roughly between component 1 
($-140$ \kms) and component 2 ($-110$ \kms).  Component 1 appears to be 
present in the low ions as well as the high ions.  However, component 2 
is extremely weak in the low ions.  The low ions may show multiple very 
weak components over the velocity range $\sim -135$ to $-90$ \kms.


\subsection{Column Density Ratios and Possible Ionization Mechanisms}
\label{sec:ratios}

Two HVC components are seen in the highly ionized \ion{C}{4} and \ion{Si}{4}, 
while there are at least 3 components seen in the low ions.  The component 
structure is less certain in the \ovi\ due both to the lower resolution of 
\fuse\ than \stis\ and intrinsically broader absorption in \ovi.  Determining 
the mechanism(s) by which these species are ionized may aid in constraining 
the environment in which these HVCs originate.

The principal diagnostics of the ionization mechanisms are the column density 
ratios of various ions; these can be compared with other sight lines and also 
with predictions from theoretical models.  In Table~\ref{tab7} we report the 
high ion column density ratios $N$(\civ)/$N$(\siiv), $N$(\civ)/$N$(\nv), and 
$N$(\civ)/$N$(\ovi).  The ratio $N$(\civ)/$N$(\siiv) is derived from the results 
of component fitting, the ratio $N$(\civ)/$N$(\nv) is derived from the results 
of component fitting for \civ\ and the $3\sigma$ upper limit for \nv\ derived 
from the AOD method, and the ratio $N$(\civ)/$N$(\ovi) is derived from the 
ratio of their $N_a(v)$ profiles.  Figure~\ref{f10} shows the ratios of apparent 
column densities for various ions as a function of \vlsr.  For the ratios 
involving \ovi\ (\fuse) we smoothed and rebinned the \stis\ data (\siiv\ and 
\civ) to match the resolution and pixel size of \fuse. 

\begin{deluxetable*}{lcccc}
\tabletypesize{\footnotesize}
\tablecolumns{5}
\tablewidth{0pt}
\tablecaption{Component-to-Component Column Density Ratios\tablenotemark{a}
\label{tab7}}
\tablehead{
\colhead{Component Number} & 
\colhead{$V_{LSR}$ [km s$^{-1}$]} &
\colhead{$N$(\ion{C}{4})/$N$(\ion{Si}{4})} &
\colhead{$N$(\ion{C}{4})/$N$(\ion{N}{5})} &
\colhead{$N$(\ion{C}{4})/$N$(\ion{O}{6})\tablenotemark{b}} 
}
\startdata
1 & $\sim-142$ & $4.2\pm0.7$ & $>3.3$ & $>2.1$ \\
2 & $\sim-111$ & $11.5\pm2.3$ & $>6.0$ & $0.4 - 3.2$ \\
\cutinhead{Predicted Ratios for Different Ionization Mechanisms\tablenotemark{c}}
Radiative Cooling\tablenotemark & & $8.7 - 33.0$ & $0.9 - 2.7$ & $0.0 - 0.2$ \\
Radiative Cooling\tablenotemark{d} & & $3.0 - 40.0$ & $0.5 - 13.0$ & $0.1 - 5.0$ \\
Conductive Interfaces\tablenotemark{e} & & $5.5 - 30.9$ & $0.8 - 2.6$ & $0.1 - 0.9$ \\
Turbulent Mixing Layers\tablenotemark{f} & & $0.9 - 28.8$ & $7.4 - 29.5$ & $1.1 - 7.4$ \\
Shock Ionization\tablenotemark{g} & & $1.3 - 33.0$ & $0.7 - 38.6$ & $0.0 - 1.1$ \\
Supernova Remnant (SNR)\tablenotemark{h} & & $8.6 - 16.7$ & $2.4 - 2.9$ & $0.1 - 1.9$ \\
Halo SNR\tablenotemark{i} & & \nodata & $1.8 - 9.0$ & $0.1 - 3.0$ \\
\enddata
\tablenotetext{a}{The values adopted for \ion{Si}{4} and \ion{C}{4} are from 
Table~\ref{tab5}.  The values for \ion{O}{6} and \ion{N}{5} are taken from 
Table~\ref{tab3} with \ion{N}{5} being $3\sigma$ upper limits.}
\tablenotetext{b}{The ratios for $N$(\ion{C}{4})/$N$(\ion{O}{6}) are derived 
from Figure~\ref{f10} (see text for more
details).
 We quote a lower limit for component 1 and a range for component 2.}
\tablenotetext{c}{The references for the collisional ionization mechanisms are: 
radiative cooling: Heckman et al. (2002), Gnat \& Sternberg (2007), Edgar \& 
Chavelier (1986), conductive interfaces: Borkowski et al. (1990), turbulent
mixing layers: Slavin et al. (1993), shock ionization: Dopita \& Sutherland (1996), 
SNRs: Slavin \& Cox (1992), halo-SNRs: Shelton (1998).}
\tablenotetext{d}{The quoted ratios assume gas cooling from $10^6$ K with a cooling 
flow velocity of $100$ \kms\ for isobaric and isochoric (and intermediate) conditions.  
The data was taken from Edgar \& Chevalier (1986) for the first
set of values, and from Gnat \& Sternberg (2007) for the second.}
\tablenotetext{e}{The quoted ratios assume magnetic field orientations in the range 
$0-85^\circ$, and interface ages in the range $10^5-10^7$ yrs.  These ratios should 
be considered crude estimates as they were estimated from graphs in
Borkowski et al. (1990).}
\tablenotetext{f}{The quoted ratios assume gas with entrainment velocities in the 
range $25 - 100$ \kms\ and mixing-layer temperatures in the range $1 - 3 \times 10^5$ K.}
\tablenotetext{g}{The quoted ratios assume shock velocities in the range $150 - 500$ 
\kms, and magnetic parameters in the range $0 - 4$ $\mu$G cm$^{-3/2}$.}
\tablenotetext{h}{The quoted ratios are for SNR ages $10^{5.6-6.7}$ yr.}
\tablenotetext{i}{The quoted ratios are for SNR ages $10^{6.0-7.2}$ yr.}
\end{deluxetable*}

Component 1 ($-162$ to $-126$ \kms) is characterized by strong absorption in 
the low ions and weak absorption in \ovi.  From Figure~\ref{f10}, we place 
a limit log($N$(\civ)/$N$(\ovi)) $\gtrsim +0.5$ ($3\sigma$).  If one integrates 
the \ovi\ over the full velocity range of $-162$ to $-126$ \kms, this ratio 
is log($N$(\civ)/$N$(\ovi))$=+0.3\pm0.1$.  However, the straight integration 
is dubious since the component structure in \civ\ appears to be different 
than that of \ovi.  The \civ\ to \siiv\ ratio seen in Figure~\ref{f10} is 
log($N$(\civ)/$N$(\siiv)) $\sim +0.6 \pm 0.2$, consistent with the results 
of component fitting.  Both \siii\ and \siiii\ are more abundant than 
\siiv\ in this component.  We find log($N$(\siii)/$N$(\siiv))$\approx+0.7$ 
for component 1.

The low ions (e.g., \oi, \siii) show a component centered between components 1 
and 2 of the high ions at \vlsr$\sim-127$ \kms.  There is less or little 
absorption in \ovi, \civ, and \siiv\ at these velocities.  The \civ\ and 
\siiv\ absorption in this region is consistent with the contributions from 
the wings of components 1 and 2.  The \oi/\siii\ ratio in this range is 
about twice as high as in component 1, suggesting a slightly smaller ionization 
fraction, $x(H^+)\sim0.9$.

Component 2 ($-126$ to $-90$ \kms) is characterized by strong \ovi\ absorption 
and weak absorption for the low ions.  From Figure~\ref{f10}, we find 
$+0.4 \lesssim$ log($N$(\ion{C}{4})/$N$(\ion{O}{6})) $\lesssim +3.2$.  Integrating the 
\ovi\ over the full velocity range of $-126$ to $-80$ \kms\ gives 
log($N$(\ion{C}{4})/$N$(\ion{O}{6})) $=+0.1\pm0.1$.  It is unclear which 
is more appropriate because the \ovi\ and \civ\ profiles do not share the same 
component structure in this velocity range, however, they could be part 
of the same physical structure.  The \civ\ to \siiv\ ratio seen 
in Figure~\ref{f10} is log($N$(\ion{C}{4})/$N$(\ion{Si}{4})) $\sim +1.0 \pm 0.3$, 
consistent with the results of component fitting.  Both \siii\ and \siiii\ 
are also more abundant than \siiv\ in this component like in the other one, 
although the \siii\ to \siiv\ ratio decreases toward higher velocities.

\ovi\ dominates the region from $-100$ to $-80$ \kms\ while the \siiv\ and 
\civ\ absorption continues to decline, and there is very little if any 
absorption in the low ions.  This shows that several ionization mechanisms 
are likely at play.

To investigate the possible mechanisms responsible for the production of the 
highly-ionized high velocity gas, we focus on \siiv, \civ, \nv, and \ovi.  We 
compare the ratios of the high ions with those predicted from selected 
theoretical models in Table~\ref{tab7}.  Although we compare our results 
with all these models, we only discuss the models for which all the ratios 
are consistent with our measured values.

The ratios predicted from the models assume solar relative abundances.  Where 
the models used older estimates of the solar abundance, we have adjusted the 
results to current solar abundance estimates adopted from Asplund et al. (2005).  
The adjustments are done following Fox et al. (2004): 
$\log[N(X)/N(Y)]^\sun_{\rm{new}}=\log[N(X)/N(Y)]^\sun_{\rm{old}}+\Delta\log A^\sun_X-\Delta\log A^\sun_Y$ where $\Delta\log A^\sun_X\equiv\log A^\sun_X(\rm{adopted})-\log A^\sun_X(\mbox{old})$.  
It would however, be preferable to recalculate the models with updated atomic 
parameters and solar abundances.  

In general, the higher stages of ionization become more important as the LSR 
velocity increases from $-150$ to $-90$ \kms.  If collisional ionization 
dominates the ionization of the high ions, the bottom two panels of 
Figure~\ref{f10} suggest component 2 is at a higher temperature than 
component 1.  If the gas of component 2 were in collisional ionization 
equilibrium (CIE), it would require temperatures of ($1-2$)$\times 10^5$ K 
to match the \civ/\ovi\ ratios (Gnat \& Sternberg 2007).  However, the limits 
on $N(\mbox{\nv})$ and the strength of \siiii\ are inconsistent with pure CIE.

Of the possible mechanisms that may give rise to the high ions for component 1, 
only non-equilibrium radiative cooling (RC; Gnat \& Sternberg 2007), turbulent 
mixing layers (TMLs; Slavin et al. 1993, Esquivel et al. 2006), and halo-supernova 
remnants (halo-SNRs; Shelton 1998) give column density ratios consistent with 
the observations (see Table~\ref{tab7} and references therein).  In the case 
of radiative cooling, only near the temperature $2\times 10^4$ K do the predicted 
column densities agree with the observed column densities and ratios.  This 
temperature is allowed by the $b$-values of this component (see Table~\ref{tab5}).  
However, the RC model predicts a higher \siii\ to \siiv\ ratio 
($\log{\mbox{N(\ion{Si}{2})/N(\ion{Si}{4})}}\approx 0.8-1.0$) than observed 
in this temperature regime 
($\log{\mbox{N(\ion{Si}{2})/N(\ion{Si}{4})}}\approx 0.2-0.8$).  
Therefore, the RC model cannot single-handedly explain the observations.  
The TML model of Slavin et al. for $\log{T}=5.3$ and $v=100$ \kms, where $T$ 
is the temperature of the mixing layer and $v$ is the transverse velocity, 
best matches the observed \civ/\siiv\ ratio, although our value of $4.17\pm0.65$ 
is still significantly higher than the metallicity-corrected value this model 
predicts.\footnote{The metallicity corrected values are 0.87--2.51, corresponding 
to velocities of 25 and 100 \kms, respectively.  Our ratio of $4.17\pm0.65$ is 
within the preadjusted range of 1.58--4.57 based on the abundances of Grevesse 
(1984).}  This temperature is also above the maximum temperature allowed from 
component fitting for this component.  A velocity higher than the highest they 
considered (100 \kms) may improve the agreement given the trends in their predictions.  
Our \civ/\siiv\ ratio for component 1 is consistent with the TML models of 
Esquivel et al. (2006).  Esquivel et al. assume ionization equilibrium and their 
models are evolving, having not reached a steady state.\footnote{The models of 
Slavin et al. do not assume ionization equilibrium.  They differ from Esquivel 
et al. in that they predict the column densities assuming a steady state has been 
reached.}  The \siiv\ column density was not calculated by Shelton (1998) for her 
halo-SNR model, so we are unable to test our most stringent ratio against the 
halo-SNR model for either component.

For component 2, TMLs, shock ionization (SI; Dopita \& Sutherland 1996), and 
halo-SNRs give column density ratios consistent with our measurements.  The RC 
models of Gnat \& Sternberg (2007) can match the observed ratios of component 2.  
However, the total column densities predicted by the models do not match the 
observed column densities where the ratios are consistent.  The observed 
\civ/\siiv\ ratio fits best with the TML model of Slavin et al. for 
$\log{T}=5.0$ or 5.5 and $v=25$ \kms.  The highest of these temperatures 
is inconsistent with the \civ\ $b$-value.  The TML model of Esquivel et al., 
on the other hand, seems to be incompatible with our \civ/\siiv\ ratio.  A 
better agreement may be possible at a lower transverse velocity than the 
minimum velocity they consider (50 \kms).  The SI model, which matches the 
observed \civ/\siiv\ ratio of $11.5\pm2.3$, has a magnetic parameter of 
$2\mu$G cm$^{3/2}$ and a shock velocity of 200 \kms.  This model could 
conceivably give similar ratios for the same magnetic parameter but with 
a velocity somewhere between 300 and 400 \kms, given the trends in their results. 

The multiphase nature of this gas combined with its complicated component 
structure makes determining the mechanism(s) responsible for the ionization 
difficult.  The \zngi\ sight line is most likely too complicated to be 
described by a single model, nor is it likely that the idealized situations 
in which the models are run encompass the true nature of this gas.  We 
cannot rule out turbulent mixing layers or halo supernova remnants for the 
$-140$ \kms\ cloud (component 1); similarly, we cannot rule out turbulent 
mixing layers, shock ionization, or halo supernova remnants for the $-110$ 
\kms\ cloud (component 2) as possible ionization mechanisms.  The kinematics 
of the HVCs toward \zngi\ are not inconsistent with these mechanisms.

Given the small $b$-values for component 1, it may be likely that 
photoionization also plays an important role in ionizing that component.  
Although we will show that \zngi\ itself is unlikely to provide for its 
ionization, the gas along the \zngi\ sight line must be subject to radiation 
escaping from the Milky Way and also the extragalactic background radiation 
(e.g., see \S~6 of Fox et al. 2005).  Furthermore, radiation emitted from 
cooling hot gas could also play a role if these HVCs are in close proximity 
to such material.  Photoionization could explain the presence of \cii, \siii, 
\siiii, and the other low ions, and Knauth et al. (2003) showed that it is 
possible to match our observed \civ/\siiv\ ratio via photoionization by 
X-ray/EUV emitting gas.  However, neither the Milky Way's escaping radiation 
nor the extragalactic background radiation can account for the observed \ovi\ 
(Fox et al. 2005).  The ionization of the high ions, and \ovi\ in particular, 
must involve collisions.  Moreover, since the high and low ions seem to 
coexist in component 1, and possibly throughout parts of component 2, there 
must be changing physical conditions throughout this multiphase gas.

We finally note that $N(\mbox{\sthree}) > 2 \times N(\mbox{\stwo})$ in 
component 1, as well  
 as integrated over the HVCs (see Table \ref{tab4}).  This is unusual for gas  
 in the Milky Way.  It is in contrast with the ionization state of  
 the photoionized warm ionized medium of the Galaxy for which \stwo\  
 makes up $\sim75\%$ of the sulfur (as demonstrated by Haffner et  
 al. 1999 and the data given in Howk et al. 2006).  This also  
 suggests a relatively high state of ionization for this gas, even  
 for the component that seems likely to be affected by  
 photoionization.  This may be reflecting the presence of hot  
 collisionally-ionized material or of a strong, hard ionizing  
 spectrum (but not so hard that is overionizes Si since in component  
 1 $N(\mbox{\siiv}) < N(\mbox{\siiii})$).  Helium recombination radiation 
could play a role in ionizing \stwo\ to \sthree.



\section{The Circumstellar Hypothesis}
\label{sec:photoionizationmodel}

Here we examine the hypothesis that the HVCs seen toward \zngi\
represents circumstellar material near  \zngi\ itself.  As a PAGB
star, \zngi\ may be expected to have some circumstellar material.
Although \zngi\ is sufficiently hot to provide for the ionization of a
planetary nebula (PN), no H$\alpha$ emission is detected about this
star (Napiwotzki \& Heber 1997).  PAGB stars in globular clusters
seem less likely to give rise to a visible planetary nebula phase than
their field star brethren (Moehler 2001).

The high-velocity gas along this sight line is significantly different
than planetary nebula gas.  The velocity of the HVCs relative to the
stellar photosphere is $\sim-150$ to $-190$ \kms, much higher than
planetary nebulae studied in absorption (e.g., K648 in M 15 has an
expansion velocity of 12--17 \kms\ according to Bianchi et al. 2001,
while the young PN ESO~457--2 studied by Sterling et al. 2005 shows an
expansion of only $\sim30$ \kms).  Furthermore, the columns of
material in PNe are typically significantly larger than those seen
here (e.g., Williams et al. 2003, Sterling et al. 2005), with many
showing very strong absorption from high ions (e.g., \civ) or from
excited fine structure states (e.g., \ion{S}{3}$^*$, \ion{O}{1}$^*$,
\ion{Si}{2}$^*$; see Sterling et al. 2005, Williams et al. 2003,
2008).  The latter are a reflection of the relatively high densities
of PN shells, with values $n_e > 1000$ \percc\ being the norm
(Williams et al. 2003, Sterling et al. 2005, 2007).  These densities
are in strong contrast to the upper limit of $n_e \le
(0.36\,\mbox{\percc}) T_4^{1/2}$ reported in
Section~\ref{sec:physicalconditionsabundances}.

While the high velocity gas seen toward \zngi\ seems unlikely to
represent a traditional planetary nebula about the star, it could
still be circumstellar in nature.  In this scenario, the density of
the material would be too low and/or the material too close to the
star to produce detectable \halpha\ emission.  However, the lower
limits to the path lengths derived in Section~\ref{sec:pathlength} are
inconsistent with this.  In addition, the lack of \cii* is
inconsistent with gas very near the star, where radiative pumping can
play a role in populating the upper $^2P_{3/2}$ fine structure level
of \cii.  For example, Trapero et al. (1996) observed circumstellar
HVC gas (at LSR velocities of $-158$ and $-51$ \kms) associated with
the close binary $\eta$ Tau.  They observed \cii* $\lambda1335$
absorption which was much stronger than the \cii\ $\lambda1334$
absorption in the spectra of both stars of the binary, with optical
pumping from $\eta$ Tau responsible for the strong \cii*.

Nonetheless, in order to strongly rule out the circumstellar
hypothesis we also investigate photoionization models of material
about the star \zngi.  Any circumstellar material is going to be
strongly affected by the radiation from the central star.  We should
be able to produce models consistent with the observations of the HVCs
if this is the case.  The results of these models strongly suggest
this hypothesis is not viable.

We use the Cloudy ionization code (version 07.02.00; last described by
Ferland et al. 1998) to calculate the ionization structure of gas
about \zngi.  The ionizing spectrum is assumed to be a model stellar
atmosphere calculated with TLUSTY (Hubeny \& Lanz 1995).  We assumed
an effective temperature $T_{\rm eff}=45,000$ K and a surface gravity
$\log g=4.48$ with a luminosity $\log L/L_\odot = 3.5$ (see Table~1).
The elements H, He, C, and N were treated in non-LTE, while other
elements were treated in LTE.  We assume the atmosphere is made up of
99\% He (by number), with C and N at 10 times the solar abundance 
(by total mass), and
O at twice solar.  All other elements were scaled to solar abundances
except Fe, for which we adopt [Fe/H]$\,=-1.27$.  Shocks within the
wind of \zngi\ might produce X-rays, and we have tested the effects of
including wind-produced X-rays.  For this, we assume emission from a
$T=10^6$ K plasma with a luminosity relative to the stellar bolometric
luminosity of $L_X / L_{\rm{bol}}=10^{-7}$ (e.g., Pallavicini et
al. 1981).  The inclusion of X-rays has very little impact on our
models.

We assume the gas is distributed in a constant density shell with $
n_{\rm H} = 0.4$ \percc.  This is equivalent to the 3$\sigma$ limit to
the electron density from the integrated \ion{C}{2}$^*$ diagnostic
(the photoionized gas from the models has a temperature $T\approx10^4$
K) or a 2$\sigma$ limit for component 1 alone 
(see \S~\ref{sec:electrondensity}).  The gas is nearly fully ionized, so
$n_{\rm H} \approx n_e$.  We adopt a gas-phase metallicity of $\log
Z/Z_\odot=+0.20$ given our [O/H] measurement, and assume solar
relative abundances.  (The relative solar abundances of the gas 
should not be compared with those of the photosphere, which has 
negligible hydrogen.  The HVCs, if circumstellar, would represent 
gas lost from a much different part of the star than as seen as 
the current photosphere.)  We do not include dust for heating or
attenuation purposes, but we do include the effects of a Galactic
cosmic ray background.  We calculate the ionization structure of the
gas from an initial inner shell radius, proceeding until the \HI\
column of the HVC is matched.  We calculate the total column densities
of ions in the gas for inner shell radii in the range $\log (r_0/{\rm
  cm}) = 16.0$ to 20.0, i.e., 0.003 to 33 pc.  The lower value is
similar to the inner radii of some planetary nebulae (e.g., Sterling
et al. 2007).  The upper value is unphysically large for such a
structure, but we extend the calculations that far to probe the
sensitivities of the column densities to this parameter.

\begin{figure}
\epsscale{1.15} 
\plotone{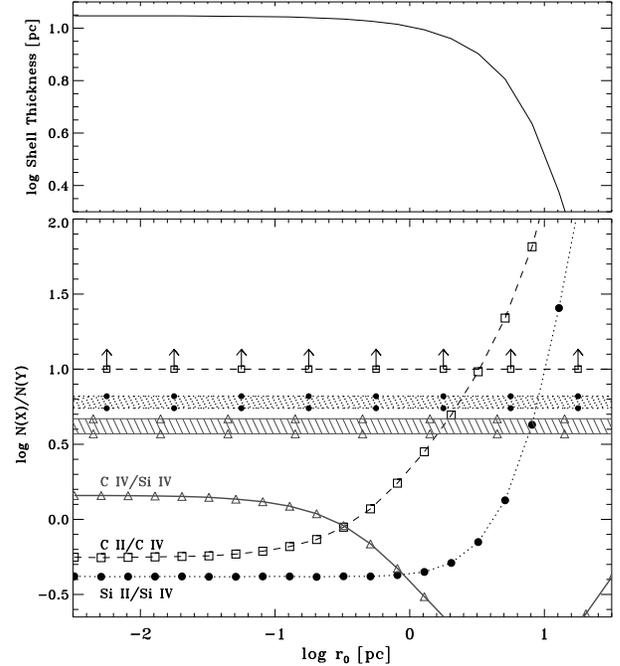}
\caption{Cloudy models of a circumstellar shell about ZNG 1.  {\em
    Top:} The shell thickness as a function of inner shell radius,
  $r_0$.  This depends on the assumed hydrogen density and the
  ionizing flux of the star.  We assume $n_{\rm H}=0.4$ \percc, near
  the upper limit provided by the \ion{C}{2}$^*$; lowering the density
  would increase the shell thickness.  {\em Bottom:} The column
  density ratios as a function of inner shell radius.  We plot the
  model predictions for the \ion{Si}{2}/\siiv, \ion{C}{2}/\civ, and
  \civ /\siiv\ ratios as the smooth curved lines.  The observational
  constraints for component 1 are shown using the same symbols.  The
  lower limit is for \ion{C}{2}/\civ\ given the likely saturation in
  the \ion{C}{2} profile.  The $\pm1\sigma$ range is shown for the
  other two ratios.  The \civ /\siiv\ ratio (triangles) cannot be
  matched by our models.  The \ion{Si}{2}/\siiv\ ratio is matched for
  an inner shell radius of $\approx9$ pc, which is consistent with
  the constraint provided by \ion{C}{2}/\civ.  As discussed in the
  text, the \civ /\siiv\ ratio for the integrated HVC is even further
  from the models.
  \label{f11}}
\end{figure}

The results of the photoionization models are inconsistent with a
circumstellar origin for the HVC toward \zngi\ for a number of
reasons.  Similar to Section~\ref{sec:pathlength}, the most
fundamental difficulty is the predicted thickness of the resulting
shell, which is shown in the top panel of Figure \ref{f11}.
Because the limits on the density are so low, a large thickness is
required to match the observed \HI\ column density of the HVC.  For
all models with inner shell radii $r_0 < 1$ pc, the thickness of the
shell exceeds 10 pc, implying a shell diameter in excess of the core
radius and the half-mass radius of the globular cluster M5 itself
(Harris 1996).  It is highly unlikely that a shell of such thickness
and low density could survive in this environment.  Furthermore, such
sizes are well in excess of typical PN size scales, as well as typical
sizes for circumstellar material observed about typical PAGB stars
(see e.g., discussion in Sahai et al. 2007).  To find shells with
thicknesses $<2$ pc requires $r_0 \gtrsim 20$ pc; once again, such a
large inner surface to the shell is inconsistent with a circumstellar
origin.

We stress that the derived thicknesses are robust predictions of the
models dependent only on the adopted density, the \HI\ column density,
and the ionizing luminosity of the star.  The low density and large
luminosity of the star drive the density of {\em neutral} hydrogen to
very low values, thereby requiring a large path length for even the
very low \HI\ column of this HVC.  The predicted thickness is
insensitive to the presence of X-rays from the wind of \zngi.  This
calculation should also be insensitive to the details of the model
stellar atmosphere, so long as the ionizing hydrogen flux is
approximately correct (we have tested this sensitivity by calculating
models that make use of main sequence OB star atmospheres with similar
effective temperatures and surface gravities, finding very similar
results).  The thickness is, of course, sensitive to the assumed
density of the gas.  Raising the density to as high as $n_{\rm H}
\approx 1.5$ \percc\ seems difficult given the constraints from the
\ion{C}{2}* absorption and the $b$-values for component 1 (see
Table~4), and that would only decrease the shell thicknesses in
Figure~\ref{f11} by a factor of about four.

We also note that the ionic ratios found in the HVC, either from the
total integrated absorption or from the narrow component 1, are
difficult to match with our photoionization models.  This is shown in 
the bottom panel of Figure \ref{f11}.  The observed \civ
/\siiv\ ratio cannot be matched by our photoionization models.  The
maximum \civ /\siiv\ ratio produced in our models is 1.4, which occurs
for $r_0 \la 0.1$ pc.  This should be compared with the observed ratio
of $4.2\pm0.7$ (Table~6) for component 1, which has a lower $b$-value
and displays more prominent low ion absorption than component 2,
suggesting it is more likely to be photoionized.  The observed \civ
/\siiv\ ratio for the integrated HVCs (components 1 and 2) is
$7.1\pm1.0$.  The \ion{C}{2}/\civ\ and \ion{Si}{2}/\siiv\ ratios, 
which are independent of the assumed relative abundance, can
both be matched at large values of the inner radius, $r_0$.  The limit
to the \ion{C}{2}/\civ\ ratio for component 1 is matched for $r_0 \ga
3$ pc ($\ga1.5$ pc for the integrated ratio), while the
\ion{Si}{2}/\siiv\ ratio is matched for $r_0 \approx 9$ pc for both
the component 1 and the integrated ratios.  The total column densities
of \civ\ and \siiv\ predicted by the models are not consistent with
the data at the same values of $r_0$ required to match their ratio.
Negligible \ovi\ is produced via photoionization.

The difficulty in matching the \civ /\siiv\ ratio and the large inner
shell radii required by the other ionic ratios suggest that the HVC is
not ionized directly by \zngi.  This is, of course, dependent on the
shape of the stellar spectrum over the energy range 33 eV to $\ga65$
eV.  We have tested the effects of adopting alternative model 
atmospheres.  CoStar model atmospheres
(Schaerer \& de Koter 1997) of main sequence stars with similar
temperatures can match the \civ /\siiv\ ratio observed
in these HVCs.  However, they cannot simultaneously match this ratio
and the \ion{C}{2}/\civ\ and \ion{Si}{2}/\siiv\ ratios, and they too
require large path lengths inconsistent with a circumstellar origin.
We also note that Smith et al. (2002) have argued the CoStar models
produce spectra that are too hard over the 41 to 54 eV range, which
could significantly affect the \civ /\siiv\ diagnostic.  While models
invoking shock ionization in addition to photoionization by \zngi\
could be considered, these will tend to drive down the ionization
fraction of \HI, requiring even thicker shells.  In short, we feel a
robust conclusion can be drawn from our models and the physical
conditions derived above: the properties of the HVCs toward \zngi\ are
inconsistent with the hypothesis that these clouds have a
circumstellar origin.


\section{Discussion}
\label{sec:discussion}


Along the \zngi\ sight line, high-velocity absorption is seen in \ovi,
\civ, and \siiv\ along with lower ionization and neutral states. As we
discussed previously, neither a circumstellar origin (see
\S~\ref{sec:photoionizationmodel}) nor a local origin (e.g., in Loop I;
see \S~\ref{sec:sightline}) is likely for this absorption. The
high-velocity absorption toward \zngi\ therefore represents an example
of interstellar highly-ionized HVCs, sharing ionization and other
characteristics (e.g., multiphase structure, column density ratios, 
$b$-values) with highly-ionized HVCs observed toward extragalactic
objects (e.g., Fox et al. 2006). The presence of the high ions,
especially \ovi, suggests the presence of hot gas ($10^6$ K)
interacting with cooler gas ($10^4$ K) in the regions giving rise to
this absorption.  The theoretical models capable of matching the 
high-ion column densities and column density ratios in these HVCs include
turbulent mixing layers, shocks, and supernova remnants.  Furthermore, the
multiphase nature of this gas implies that more than one ionization
mechanism is responsible for the creation of the ions in the HVCs.
While collisional ionization is required to explain the \ovi\ and
portions of the other high ions, photoionization is likely to play an
important role in the lower ionization species.

The \ovi\ absorption toward \zngi\ is reminiscent of high-velocity
``wings'' seen in \ovi, \ion{C}{3}, and \ion{H}{1} toward AGNs (e.g.,
Sembach et al. 2001; 2003, Savage et al. 2005, Fox et al. 2006).  
These authors do not identify any negative-velocity wings.  One
interpretation advanced to explain the highly-ionized wings and
other HVCs is that they probe a hot Galactic fountain or Galactic
wind. In this scenario, supernovae and stellar winds create large
overpressurized bubbles that may eject material and energy from the disk into the halo
of the Milky Way. Depending on the speeds of the outflow (often at the
sound speed of the hot gas), some material can fall back on the disk
(this is a fountain scenario, see Shapiro \& Field 1976. Houck \& Bregman 1990).
If the speeds of the outflow are greater than the escape velocity, a
galactic wind will push the material beyond the potential well of the
Galaxy. Galactic outflows (and the returning material) often have a
complex multiphase mixture of cool and hot gas, as revealed by
observations of starburst and dwarf galaxies and by galaxy wind models
(e.g., see Veilleux et al. 2005, Heckman et al. 2002, Lehner \& Howk
2007).  The HVCs along the \zngi\ sight line are complex multiphase
structures, consistent with these observations and models, and the
presence of high ions suggests they are closely connected to energetic
phenomena in the Galaxy.  The blueshifted velocities of the HVCs
toward \zngi\ indicate flows (radial or vertical) with a component
directed toward the sun.  This may be indicative of a large-scale
Galactic circulation, where we are seeing the return of material
rather than the ejection.

The origins of highly-ionized HVCs in a galactic fountain-type flow is
certainly plausible, but it requires that the HVCs originate in the
Galaxy.  Thus, their metallicities should be near solar and their
distances not too great in such a model.  The distances of
highly-ionized HVCs toward QSOs and AGNs are unknown, and the
metallicity estimates for these HVCs often depend sensitively on the
specific ionization model adopted (see below). However, the distance
and metallicity of the highly-ionized HVCs toward \zngi\ are known. The
distance of M5 is well determined and places a strict upper limit
for the distance to the HVCs of $d< 7.5$ kpc. The metallicity of the
HVC gas is [O/H] $=+0.22\pm0.10$.  Since
our metallicity estimate was derived from a comparison of \oi\ and
\HI, no ionization correction is needed (see \S~\ref{sec:abundance}).  
Together, the metallicity
and distance constraints rule out an extragalactic origin for these
HVCs; instead, it seems probable that these clouds are related to
energetic phenomena that drive gas flows in the Milky Way. 
These are the only highly-ionized HVCs confirmed to be
within the Galaxy.

The super-solar metallicity suggests these HVCs originated near the
central regions of the Galaxy.  Smartt \& Rolleston (1997) and
Rolleston et al. (2000) showed that the oxygen abundance of early
B-type main-sequence stars decreases by $-0.07 \pm 0.01$ dex kpc$^{-1}$ 
as one moves away from the Galactic center.  The expected
metallicity of disk gas at a radial distance similar to \zngi\ is thus
[O/H] $\sim+0.35$ if we calculate this from the solar neighborhood
toward the Galactic center.  The metallicity of these HVCs may be
less because they originated closer to the sun than M5.  On the other hand, 
if O is depleted by about 0.1 dex, [O/H] of the HVCs toward \zngi\ would match 
the O abundance near the Galactic center.

The close proximity of \zngi\ to the Galactic center, the complex
multiphase structure of the high-velocity gas in this direction, its 
super-solar metallicity, and
its large velocities raise the possibility that these HVCs may not
only be linked to a Galactic fountain, but in some way to a wind driven
from the vicinity of the Galactic center (e.g., Sofue 2000, Yao \&
Wang 2007).  Bland-Hawthorn \& Cohen (2003) and Keeney et al. (2006)
each produced a phenomenological model of a Galactic nuclear wind. The
former produced their model based on {\it Midcourse Space Experiment}
observations of infrared dust emission from the Galactic center and
{\it ROSAT} X-ray observations, while the model of Keeney et al. was
developed to explain highly-ionized HVCs along the Mrk~1383 and
PKS~2005--489 sight lines.  The model presented by Bland-Hawthorn \&
Cohen has a bipolar wind with initial velocities of $1700-3000$ \kms,
rising from the Galactic center in a $\sim45^\circ$ conical shape,
which becomes cylindrical at a height above the plane of $\sim 5$ kpc
and with a radius of $\sim6$ kpc.  The model described by Keeney et
al. is similar to that of Bland-Hawthorn \& Cohen, except that the
conical shape turns cylindrical at a $z$-height of $\sim2$ kpc and the
cylinder has a radius of $\sim1.5$ kpc.  \zngi\ lies within the outer
bounds of the outflow cone in the wind model of Bland-Hawthorn \&
Cohen and just outside that of Keeney et al.  While the initial velocities 
at the base of such a wind 
are high, Sofue (1984) showed that this wind could slow down considerably.  
The shock velocities of an evolving wind considered by Bland-Hawthorn \& 
Cohen (2003) and Sofue (2000, 1984) are of the order of one to 
a few hundred \kms.  If the model of
Bland-Hawthorn \& Cohen is representative of the Milky Way's nuclear
wind, then the highly-ionized HVCs observed toward \zngi\ may be gas
near the leading edge of the conical section of the wind that is
falling back onto the Galaxy.  Such infall can occur if thermal
instabilities form denser gas with too little buoyancy to be supported
against gravity.

If a significant amount of material is participating in outflows or
fountain-like processes in the inner Galaxy, other sight lines passing
through this region should show signs of these flows.  A few
extragalactic sight lines that are within about $\pm25^\circ$ longitude
from the Galactic center were observed with {\em FUSE}\ or STIS, and
all of them (with enough signal) show highly-ionized high-velocity
absorption.  These include \zngi\ (\vlsr\ $<0$), PG~1553+113 (\vlsr\
$<0$) (see below), Mrk~1383 (\vlsr\ $>0$) (Keeney et al. 2006, Fox et
al. 2006), PKS~2005--489 (2 with \vlsr\ $<0$) (Keeney et al. 2006, Fox
et al. 2006), ESO~141--G55 (\vlsr\ $>0$) (Fox et al. 2006), and
PKS~2155--304 (\vlsr\ $>0$) (Collins et al. 2004, Fox et al. 2006). Similar to
the HVCs toward \zngi, all these HVCs are multiphase.  Therefore,
sight lines that pass through or near the central region show the signature
of outflow and infalling material, consistent with large circulation
motion with the inner regions of the Milky Way.  We note that the
sight line toward PG~1553+113 ($l=21\fdg9$, $b=+44\fdg0$) lies only
$13\fdg1$ from \zngi\ and shows a high negative velocity \ovi\
absorption detected at 3.5$\sigma$ (see \S~\ref{sec:sightline}).  
The velocity ($-170 \leq v_{\rm LSR} \leq -100$ \kms), column density 
($\log N(\mbox{\ovi})=13.57^{+0.11}_{-0.13}$), and the shape of the \ovi\ 
profile (see Figure~6 in Fox et al. 2006) are quite similar to those 
of the HVCs toward \zngi, suggesting that this HVC could be part of 
a similar flow.

Unfortunately, the metallicity for the HVC toward PG~1553+113 cannot
be estimated: the column density of \HI\ cannot be measured due to low
signal-to-noise ratios in the \fuse\ SiC channel data.  For the other
sight lines mentioned above, which probe the inner Galaxy, $N($\HI$)$ is only
estimated for two HVCs in the spectrum of PKS~2155--304 (Collins et
al. 2004) and estimated with very large uncertainties for one HVC in
the spectrum of PKS~2005--489 (Keeney et al. 2006). For PKS~2155--304,
we calculate [O/H]$\,<+0.37$ and $<+1.62$ for the $-140$ and $-270$
\kms\ clouds, respectively, based on the column densities of
\oi\ and \HI\ estimated by Collins et al. (2004). These values {\it could} 
be consistent with a super-solar metallicity. On the other
hand, photoionization models by Collins et al. give subsolar values of
$-0.47^{+0.15}_{-0.24}$ and $-1.20^{+0.28}_{-0.45}$, inconsistent with
the origins of these clouds in a Galactic fountain or outflow.  We
note that these uncertainties only include errors in the column
density measurements, with no contribution from uncertainties in the
modeling.  These model-derived metallicities are highly dependent on
the adopted ionizing spectrum and model assumptions. For example, J.C. Howk
et al. (2007, in preparation) show that using a QSO-type and QSO+galaxy type
radiation fields can produce metallicities that are different by a
factor three in IGM absorbers. Collins et al. adopted a QSO-only radiation field, which
provided a better fit to the data than a purely stellar radiation
field.  However, it is not clear how their metallicity estimates would
change if their radiation field included ionizing radiation from the
Galaxy as well as QSOs.  Furthermore, \siiv\ and \civ\ were used to
constrain their models, even though it is likely that collisionial
processes may produce part of the absorption of these high ions (see
discussion in Collins et al. 2004).  For PKS~2005--489, the metallicity
is largely unknown given the order of magnitude uncertainty in
$N($\HI$)$.  We therefore believe that without a
full understanding of the ionization in highly-ionized HVCs,
metallicity estimates derived purely from photoionization models
should be considered tentative at best.

Highly-ionized HVCs along the sight lines HE~0226--4110,
PG~0953+414 (Fox et al. 2005) and PG~1116+215 (Ganguly et al. 2005)
also have metallicities derived from a comparison of \oi\ to \HI. The
metallicities of the HVCs toward HE~0226--4110 ($+175$ \kms\ cloud),
and PG~1116+215 ($+100$ and $+184$ \kms\ clouds) are [O/H] $<-0.07$,
$<+0.05$, and $-0.66^{+0.39}_{-0.16}$, respectively.  Those values are
all $\gtrsim2\sigma$ lower than the [O/H] measurement for the HVCs
toward \zngi.  The low metallicities may imply an extragalactic origin 
for these HVCs.  This does not affect the hypothesis that the HVCs toward 
\zngi\ trace outflows or circulation/feedback in the inner Galaxy 
since these sight lines do not
pass through the central regions of the Galaxy.  It is also not clear
that these metallicities are significantly subsolar.  Like for the
\HI\ HVCs, it is very unlikely that highly-ionized HVCs have a single
origin. For instance, based on their kinematics and projection on the
sky, several highly-ionized HVCs appear to be associated with known
\HI\ HVCs such as Complexes A and C, the Magellanic Stream, and the
Outer Arm (e.g., see Fox et al. 2006). The HVCs toward \zngi\ show, for
the first time, that some of the highly-ionized HVCs likely have their
origins in the Galaxy. We note that the ionization characteristics and
physical conditions of the \zngi\ HVCs are similar to other
highly-ionized HVCs (e.g., $N(\mbox{\civ})/N(\mbox{\siiv})$, 
$N(\mbox{\civ})/N(\mbox{\ovi})$, $b$-values, multiphase structure).  
This may imply that similar ionization
mechanisms may be at work in creating the high ions, but it also means
that highly-ionized Galactic and extragalactic HVCs cannot be easily
separated based on their ionization characteristics alone. 
We finally note that the properties (i.e., no \HI\ 21 cm emission, 
$|v_{\rm LSR}| > 100$ \kms) of the highly-ionized HVCs toward \zngi\ fit 
the category of highly-ionized HVCs considered by Nicastro et al. (2003).  
These authors proposed a Local Group origin for highly-ionized 
HVCs because it would better explain their kinematic distribution.  
The highly-ionized HVCs toward the \zngi\ sight line show that this 
argument alone is rather weak.  Distances
and metallicity estimates are key to distinguish Galactic from
extragalactic HVCs, but these are most often estimated indirectly for
highly-ionized HVCs, relying on photoionization models.  More observations of 
distant stars may allow 
direct distance limits for other clouds, as we have done for
\zngi.

If a Galactic nuclear wind exists and produces significant columns of
high ions such as \siiv\ and \civ\ (that will be observable with the
future Cosmic Origins Spectrograph, COS), it should be visible along other
sight lines in the general direction of \zngi.  Stellar sight lines in
this region of the sky may be useful for placing distance constraints
on the gas, while AGNs could be used to probe its structure, including
its velocity structure.  The QSO SDSS J150556.55+034226.3,
observed with \stis, lies within $5^\circ$ of \zngi.  This QSO has
been observed with a low resolution mode of \stis, but is bright
enough for future high-resolution UV spectroscopy.  

Finally, we note that these HVCs could be related to HVC complex L 
given their relative proximity on the sky and velocities 
(see \S~\ref{sec:sightline}).  Since the angular 
separation is several times the angular extent of the \HI\ component of 
complex L, this association is, however, 
by no means firm.  \halpha\ emission may be useful in not only studying this 
connection, but 
also in tracing feedback flows in the inner Galaxy.  Haffner (2005) has mapped 
intermediate and high negative velocity gas toward complex L (at $l>0^\circ$).  
He finds pervasive emission in this region at $-95\leq \mbox{\vlsr} \leq -50$ \kms.  
\halpha\ emission in this range is seen within $2^\circ$ of \zngi\ with the WHAM survey, 
though not in the pointing nearest \zngi.  Haffner also finds extended \halpha\ emission 
from ionized gas associated with complex L ($-150 \leq \mbox{\vlsr} \leq -80$ \kms).  
Future mapping 
of this emission and UV absorption with COS of the central region of the Galaxy 
 may reveal a connection between the HVCs studied here and the 
material associated with complex L.  More importantly, such observations bear on the 
connection of all this matter to feedback-driven flows in the inner Galaxy.


\section{Summary}
\label{sec:summary}

We have presented {\em Far Ultraviolet Spectroscopic Explorer}
and Space Telescope Imaging Spectrograph
observations of the PAGB star \zngi\, which
resides in the globular cluster Messier 5, using these data to study the high-velocity 
gas along this sight line.  The major results of this work are as follows.

\begin{enumerate}
  
\item 
We have analyzed the high velocity absorption toward \zngi\ and find the presence of 
\siiv, \civ, and \ovi\ along with lower ionization species such as \cii, \siii, \nii, 
\oi, \alii, \feii, and \siiii.  These high and low ions are observed in the same 
velocity range, although they do not share the same component structure.  We have 
catalogued the column densities and $b$-values using primarily the AOD and component 
fitting methods.
  
\item 
We investigated the possibility that the gas along the \zngi\ sight line is 
circumstellar.  Limits to the electron density of the gas strongly constrain 
the path length to be $\gtrsim 0.6$ pc, inconsistent with a circumstellar 
origin.  Furthermore, detailed photoionization calculations also argue against 
a circumstellar origin based on path length and metal ion ratios.

\item
We measured the metallicity of the HVCs to be $[\rm{O/H}]=+0.22\pm0.10$ using the 
\oi\ and \HI\ absorption.  This, combined with the well determined distance to the 
globular cluster M5 where \zngi\ resides (7.5 kpc), implies that these clouds have 
a Galactic origin.  This is the first evidence that highly-ionized HVCs may be 
found near the Galactic disk.  The \zngi\ HVCs have the highest metallicity of 
any known HVCs.

\item
We argue that the \zngi\ HVCs are not associated with Loop I and reside closer to 
the Galactic center than this structure.  We argue that these HVCs may be 
associated with a Galactic nuclear wind, or Galactic fountain-like circulation 
in the inner Galaxy, where the HVCs represent gas falling back to the disk.

\item
The exact details of the physical processes responsible for the ionization of 
these HVCs are yet to be resolved, and it is likely that a complex interplay 
between several processes is at work.  These HVCs are characterized by changing 
physical conditions, as component 1 ($-110$ \kms) is likely to be significantly 
affected by photoionization consistent with the small $b$-values and the stronger 
presence of lower ionization species, while component 2 ($-140$ \kms) must be mostly 
collisionally ionized as required by the strong presence of \ovi\ and less lower 
ionization species.  The theoretical models consistent with the data for 
component 1 are turbulent mixing layers and halo-supernova remnants; the 
models consistent component 2 are turbulent mixing layers, shock interfaces, 
and halo-supernova remnants.  At least qualitatively, the complex phase 
structure is consistent with energetic circulation associated with a Galactic 
wind or outflow.

\item
It will be difficult to firmly identify a population of extragalactic  
highly-ionized HVCs (including intergalactic Local Group gas or material 
condensed from an extended Galactic Corona) without direct 
estimates or limits of distance and metallicity, 
since the kinematics and ionization properties of other highly-ionized 
HVCs observed toward extragalactic sight lines appear similar to those 
of the HVCs toward \zngi.


\end{enumerate}

\acknowledgements
This work was partially supported by National Science Foundation  
grant AST 06-07731 and by the National Aeronautics and Space  
Administration under Grant No. NNX07AG93G through the Science Mission  
Directorate.  This work is based on observations made with the NASA-CNES-CSA 
{\it Far Ultraviolet Spectroscopic Explorer} (\fuse) instrument and the NASA-ESA 
{\it Hubble Space Telescope} (\hst) instrument.  \fuse\ is operated for NASA by 
the Johns Hopkins University under NASA contract NAS5-32985, and \hst\ data is 
obtained from the Space Telescope Science Institute and is operated by the 
Association of Universities for Research in Astronomy, Inc. under NASA 
contract NAS5-26555. This research has made use of the NASA Astrophysics 
Data System Abstract Service and the SIMBAD database, operated at CDS, 
Strasbourg, France.  In addition, WVD recognizes funding from FUSE grant NNG04GC44G.


 
\end{document}